\documentclass[aps,pre,twocolumn,superscriptaddress,showpacs,floatfix]{revtex4-2}
\usepackage{amsmath}
\usepackage{amssymb}
\usepackage{amsfonts}
\usepackage{lipsum}
\usepackage[dvipdfmx]{graphicx}
\usepackage{hyperref}
\hypersetup{colorlinks=true,linkcolor=blue,citecolor=blue,urlcolor=blue}
\usepackage[normalem]{ulem}

\usepackage{bm}
\usepackage{braket}
\usepackage{mathtools}

\usepackage{tikz}
\usetikzlibrary{quantikz2}
\usetikzlibrary{backgrounds,fit,decorations.pathreplacing,calc}

\begin{document}

\title{Quasi-adiabatic thermal ensemble preparation in the thermodynamic limit}

\author{Tatsuhiko Shirai}
\email{tatsuhiko.shirai@aoni.waseda.jp}
\affiliation{Waseda Institute for Advanced Study, Waseda University, Nishi Waseda, Shinjuku-ku, Tokyo 169-0051, Japan}

\date{\today}

\begin{abstract}
    %This is a brief abstract of the manuscript, summarizing the main results and significance. Keep it concise and informative.
    %A thermal adiabatic process is a fundamental thermodynamic process in which no heat is exchanged between a system and its surroundings.
    We investigate a quasi-adiabatic thermal process for preparing finite-temperature ensembles in the thermodynamic limit.
    %The method combines gate-based operations with a quantum adiabatic protocol.
    The process gradually transforms a thermal ensemble of a noninteracting system into that of an interacting system of interest over a finite operation time, with the temperature controlled by parameters associated with the entropy of the initial state.
    %The algorithm builds on this process with the free-energy minimization.
    %This algorithm builds on a purification method, which computes the ensemble average of a local observable with polynomial computational complexity, and the principle of minimum free energy.
    %The algorithm is reduced to the conventional quantum annealing for ground-state preparation when entropy is zero.
    We analyze this process in both nonintegrable and integrable spin chains with translational invariance.
    For the nonintegrable case, numerical simulations combined with a thermodynamic argument indicate that the thermal properties of local observables are accurately reproduced with a single parameter, although the operation time increases exponentially with precision.
    %We also analyze how the error depends on the system size and operation time and demonstrate its scalability by introducing a time-averaging process.
    In contrast, for the integrable transverse-field Ising model, we analytically show that an extensive number of parameters tied to local conserved quantities is generally necessary, and the performance is affected by the presence of a quantum phase transition.
    %, and that the operation time increases linearly with system size, diverging in the thermodynamic limit.
    These results clarify the potential and limitations of the quasi-adiabatic thermal process for an ensemble preparation and highlight the role of integrability in determining its efficiency.
    %Furthermore, errors at finite operation time can be understood through the Kibble-Zurek mechanism.
    %Our results extend quantum annealing from ground-state preparation to finite-temperature ensemble preparation.
\end{abstract}

\maketitle

\section{Introduction}
%Introduce the topic, highlight the importance of the research, and outline the structure of the manuscript.

The preparation of quantum thermal ensembles is a central problem in statistical physics, with broad relevance to condensed matter physics and quantum chemistry, among others.
Classical approaches, including quantum Monte Carlo methods~\cite{suzuki1993quantum}, tensor network techniques~\cite{orus2019tensor}, and machine learning approaches~\cite{nomura2021purifying}, have provided powerful tools for simulating thermal properties.
Significant progress has been made in extending these methods to strongly interacting systems and lower temperatures, yet they remain fundamentally limited by issues such as the negative-sign problem.

Quantum computing offers a promising alternative route to this challenge.
Several approaches have been proposed, including quantum Metropolis sampling~\cite{temme2011quantum, yung2012quantum}, simulations based on Lindblad dynamics~\cite{chen2023efficient,gilyen2024quantumgeneralizationsglaubermetropolis}, and imaginary-time evolution~\cite{mcArdle2019variational,yuan2019theory,motta2020determining,coopmans2023predicting}.
However, each method faces obstacles to practical realization on near-term noisy quantum devices.
For example, quantum Metropolis sampling requires deep circuits owing to phase estimation.
Variational algorithms based on free-energy minimization~\cite{wang2021variational,consiglio2024variational} have also been explored for the near-term devices, but these schemes require entropy estimation, which is computationally hard~\cite{gheorghiu2024estimating}, and often suffer from the absence of a systematic circuit ansatz.
%imaginary-time evolution~\cite{yuan2019theory,motta2020determining, coopmans2023predicting}, and variational algorithms~\cite{}.

%More fundamentally, the exact preparation of the Gibbs states at low temperatures is believed to be as hard as ground-state preparation, which is a quantum Merlin Arthur (QMA)-complete problem~\cite{alhambra2023quantum}.

On the other hand, the universality of thermodynamics implies that exact Gibbs state construction is not required to reproduce the thermal properties of local observables, i.e., operators acting on a connected region of $O(1)$ sites.
Recent developments in nonequilibrium statistical physics, such as the concept of typical states~\cite{goldstein2006canonical,popescu2006entanglement}, suggest that thermal properties can be captured without explicitly preparing Gibbs states.
%This raises a natural question: how can the thermal properties of local observables be efficiently simulated on quantum hardware?
From this perspective, simulating a (quasi-)adiabatic thermal process~\cite{nikolai2021adiabatic, zuo2024work, irmejs2025quasi, granet2025adiabatic} emerges as a promising candidate for quantum algorithms.
In this process, a thermal ensemble of a simple system is initially prepared and then transformed into that of an interacting system of interest via a quantum adiabatic protocol executed over a finite operation time.
The ensemble average of an observable over the final state can be efficiently evaluated on a gate-based quantum device~\cite{wang2021variational,consiglio2024variational,shirai2025compressed}.
Although this approach generally does not yield the exact Gibbs state, the expectation values of local observables are expected to approximate the thermal values.
However, the performance in the thermodynamic limit remains to be clarified.
%The eigenstate thermalization hypothesis (ETH)~\cite{deutsch1991quantum,srednicki1994chaos,rigol2008thermalization,mori2018thermalization}, which is generally believed to hold in ergodic systems, provides a theoretical foundation: expectation values of local observables are smooth functions of energy density, and thus any diagonal ensemble with subextensive energy variance is indistinguishable from the thermal state at the level of local observables.

%\begin{figure}[t]
%    \centering
%    \includegraphics[width=0.45\linewidth]{adiabatic_paper.pdf}
%    \caption{
%    Schematic picture for thermal state preparation based on a thermal adiabatic process.
%    The final state $\rho_\mathrm{f}$ is optimized by the free-energy minimization with respect to variational parameters $\{\phi_j\}$ assigned to the initial state $\rho_\mathrm{i}$.
%    }
%    \label{fig:adiabatic}
%\end{figure}

In this work, we study the quasi-adiabatic thermal process for preparing thermal ensembles in the thermodynamic limit.
%Figure~\ref{fig:adiabatic} presents a schematic of this algorithm.
%This process starts from a thermal ensemble of a noninteracting system, where the parameters associated with the entropy are introduced.
%Then, the system is gradually transformed into an interacting system of interest over an operation time.
%After simulating the thermal adiabatic process over an operation time $\tau$.
%, these parameters are optimized through free energy minimization. 
%Then, the variational parameters are optimized based on the free-energy minimization.
%Since the evolution is unitary, the entropy remains analytically computable from the initial state.
%to represent an ensemble of quantum states, whereas the annelaing protocol deforms a simple non-interacting ensemble into that of the interacting systems of interest.
Unlike exact Gibbs-state preparation, our aim is to reproduce the thermal properties of local observables.
%Therefore, the number of variation parameters 
To quantify accuracy, we employ the specific relative entropy as a measure of the distance from the Gibbs state~\cite{mori2016macrostate}.
We analyze the performance of this process in both nonintegrable and integrable spin chains with translational invariance.
For the nonintegrable case, numerical results combined with a thermodynamic argument indicate that a single parameter that controls the entropy of the initial state is sufficient to reproduce the thermal properties of local observables, although the operation time increases exponentially with precision.
Through the analysis, we highlight the key role of the eigenstate thermalization hypothesis (ETH)~\cite{deutsch1991quantum,srednicki1994chaos,mori2018thermalization}, which is widely believed to hold in nonintegrable systems~\cite{rigol2008thermalization, kim2014testing, yoshizawa2018numerical}.
We also investigate the effect of time averaging on the performance.
In contrast, for the integrable case we analytically show that a fine tuning of the initial state is generally necessary, and the performance is affected by the presence of a quantum phase transition.
%Moreover, deviations due to the finite operation time can be understood through the Kibble-Zurek mechanism~\cite{zurek2005dynamics}, even at finite temperatures.

This paper is organized as follows.
Section~\ref{sec:model_method} introduces a quasi-adiabatic thermal process.
Section~\ref{sec:result} shows the results for both nonintegrable and integrable spin chains.
Section~\ref{sec:conclusion} concludes with a summary and outlook for future directions.
%To overcome the difficulty, we simulate a thermal adiabatic process on quantum computers.
%A thermal adiabatic process is a fundamental process in which no heat is exchanged between a system and its surroundings.
%We propose an algorithm to simulate this process on quantum computers by hybridizing quantum gate operations with a quantum annealing protocol.
%The gate operations allow us to describe an ensemble of quantum states by introducing parameters associated with entropy.
%The quantum annealing protocol is used to gradually transform a thermal ensemble of a non-interacting system into that of an interacting system of interest.
%This algorithm builds on~\cite{wang2021variational,consiglio2024variational}, where ensemble averages of local observables can be efficiently computed.
%The algorithm is reduced to the conventional quantum annealing for ground-state preparation when entropy is zero.

%We apply this quantum algorithm to prepare a quantum thermal equilibrium ensemble, which is a fundamental problem in statistical mechanics.
%Our objective is not to construct the Gibbs state but to accurately describe the thermodynamic properties of local observables such as magnetization density.
%The specific relative entropy gives a measure on the distinguishability of two states in terms of macroscopic observable~\cite{mori2016macrostate}.

\section{Quasi-adiabatic thermal process}\label{sec:model_method}
%We first define the thermal adiabatic process and then propose a variational quantum algorithm to simulate this process for thermal state preparation.
We consider a quasi-adiabatic thermal process~\cite{nikolai2021adiabatic, zuo2024work, irmejs2025quasi, granet2025adiabatic}.
%\subsection{Initial state preparation phase}
The process begins with $N$ independent spins, where the initial Hamiltonian is given as
\begin{equation}
    H_\mathrm{i}=-\sum_{i=1}^N h_i \hat{\sigma}_i^x,
\end{equation}
where $\hat{\sigma}_i^\alpha (\alpha \in \{x,y,z\})$ and $h_i \geq 0$ are the Pauli spin operators and the magnetic field acting on site~$i$, respectively.
The initial state is prepared as the thermal state at inverse temperature $\beta_\mathrm{i}$: 
\begin{align}
    \rho_\mathrm{i}=&\frac{e^{-\beta_\mathrm{i} \hat{H}_\mathrm{i}}}{\mathrm{Tr}[e^{-\beta_\mathrm{i} \hat{H}_\mathrm{i}}]} \nonumber\\
    =& \bigotimes_{j=1}^N (\cos^2 \phi_j \ket{+_j}\bra{+_j}+\sin^2 \phi_j \ket{-_j}\bra{-_j}),
\end{align}
where $\phi_j$ is determined from the Boltzmann weight.
In this study, we regard $\{ \phi_j \}$ as variational parameters and optimize them by minimizing free-energy density of the final state under the constraint that $\sum_{j=1}^N h_j=N$~\footnote{The constraint determines $\beta_\mathrm{i}$ as a function of $\{\phi_j\}_{j=1}^N$.}.
$\ket{+_j}$ and $\ket{-_j}$ are eigenstates of $\hat{\sigma}_j^x$ with eigenvalues $1$ and $-1$, respectively.
The entropy density for a density matrix $\rho$ is defined as the von Neumann entropy $s_N(\rho)=-\mathrm{Tr} (\rho \ln \rho)/N$, and for $\rho_\mathrm{i}$ it is given by
\begin{equation}
s_N(\rho_\mathrm{i}) = - \frac{1}{N}\sum_{j=1}^N [ \cos^2\phi_j \ln (\cos^2\phi_j) + \sin^2\phi_j \ln (\sin^2\phi_j) ].
    \label{entropy_ini}
\end{equation}

%\subsection{Unitary time evolution phase}
The system then evolves according to the following protocol.
The state at time $t$ is given by
\begin{equation}
    \rho(t)=\hat{U}_t \rho_\mathrm{i} \hat{U}_t^\dagger, \quad \hat{U}_t=\mathcal{T} e^{-i\int_0^t \hat{H}(t'/\tau) dt'},
    %\frac{d\rho(t)}{dt} = -i \left[\hat{H}\left(\frac{t}{\tau}\right), \rho(t)\right], \rho(0)= \rho_\mathrm{i},
\end{equation}
where $\cal{T}$ is the time-ordering operator and $\tau$ denotes the operation time.
Here, we set $\hbar=1$.
The time-dependent Hamiltonian is linearly interpolated between the initial and final Hamiltonians: for $s_\mathrm{a} \in [0,1]$
\begin{equation}
    \hat{H}(s_\mathrm{a})=(1-s_\mathrm{a}) \hat{H}_\mathrm{i} + s_\mathrm{a} \hat{H}_\mathrm{f},
\end{equation}
where $\hat{H}_\mathrm{f}$ is the Hamiltonian of interest.
%This protocol is based on quantum annealing~\cite{kadowaki1998quantum}.
%The interpolation function $s_\mathrm{a}(t)$ is a monotonically increasing function of time, taken here linear:
%\begin{equation}
%    s_\mathrm{a}(t)=\frac{t}{\tau}
%\end{equation}
%with $\tau$ denoting the operation time.
The final state is $\rho_\mathrm{f}(\tau)=\rho(\tau)$.
The energy density for $\rho_\mathrm{f}(\tau)$ is
\begin{equation}
    e_N(\rho_\mathrm{f}(\tau))=\frac{1}{N}\mathrm{Tr} (\hat{H}_\mathrm{f} \rho_\mathrm{f}(\tau)).
    \label{eq:energyd}
\end{equation}
Since the process is unitary, the entropy density remains constant:
\begin{equation}
    s_N(\rho_\mathrm{f}(\tau))=s_N(\rho_\mathrm{i}).
    \label{entropy_fin}
\end{equation}
The process is reduced to a quantum annealing~\cite{kadowaki1998quantum} when $\phi_j=0$ for all $j$.
We adopt this process since it can be implemented on gate-based quantum devices using rotation-$Y$ gates, controlled-$X$ gates, and the Suzuki--Trotter expansion of $\hat{U}_\tau$~\cite{wang2021variational,consiglio2024variational}.

To further enhance thermalization, we introduce a time-averaging process.
Over an additional time window $\tau_\mathrm{a}$, we define the average density matrix
\begin{equation}
    \bar{\rho}(\tau,\tau_\mathrm{a}) = \int_0^{\tau_\mathrm{a}}  \rho_\mathrm{f} (\tau,t ) \frac{dt}{\tau_\mathrm{a}},
    \label{eq:time-average}
\end{equation}
where $\rho_\mathrm{f} (\tau,t ) = e^{-\mathrm{i}\hat{H}_\mathrm{f}t} \rho_\mathrm{f} (\tau ) e^{\mathrm{i}\hat{H}_\mathrm{f}t}$.
This process reduces the free-energy density: $f_N(\bar{\rho}(\tau, \tau_\mathrm{a})) \leq f_N(\rho_\mathrm{f}(\tau))$, where $f_N(\rho)=e_N(\rho)-\beta^{-1}s_N(\rho)$.

We use the quasi-adiabatic thermal process to study the thermal properties of a quantum system governed by $\hat{H}_\mathrm{f}$ at finite temperature $\beta^{-1} > 0$.
The thermal equilibrium state is given by the Gibbs state,
\begin{equation}
    \rho_\mathrm{g}(\beta)=\frac{e^{-\beta \hat{H}_\mathrm{f}}}{\mathrm{Tr} [e^{-\beta \hat{H}_\mathrm{f}}]}.
\end{equation}

It should be emphasized that neither $\rho_\mathrm{f} (\tau)$ nor $\bar{\rho}(\tau,\tau_\mathrm{a})$ exactly reproduces the Gibbs state.
The Gibbs state is diagonal in the energy eigenbasis, with a number of parameters growing exponentially with system size.
In contrast, the final states depend on only $N$ parameters $\{ \phi_j\}_{j=1}^N$, making it a compact description.

However, the ETH~\cite{deutsch1991quantum,srednicki1994chaos,rigol2008thermalization,mori2018thermalization} implies that a single parameter that controls the value of the energy density of the final state can accurately capture thermal properties of local observables.
The ETH states that for any local observable $\hat{O}$, the diagonal matrix elements obey
\begin{equation}
    [\hat{O}]_{nn}=O_{\beta_n},
\end{equation}
where $[\cdot]_{mn}=\bra{m} \cdot \ket{n}$, $\hat{H}_\mathrm{f}\ket{n} =E_n\ket{n}$, and $O_{\beta_n}$ is the thermal value at the inverse temperature $\beta_n$, which is determined by $E_n=\mathrm{Tr}(\hat{H}_\mathrm{f} \rho_\mathrm{g}(\beta_n))$.

The indistinguishability of a state $\rho$ from the Gibbs state $\rho_\mathrm{g}(\beta)$ with respect to local observables $\hat{O}$ can be quantified by the specific relative entropy:
\begin{equation}
    s_N(\rho\|\rho_\mathrm{g}(\beta)) =\frac{1}{N} \mathrm{Tr} [\rho (\ln \rho - \ln \rho_\mathrm{g}(\beta))].
    %=\beta (f(\rho)-f(\rho_\mathrm{g})).
\end{equation}
This quantity is related to the free-energy density via $s_N(\rho\|\rho_\mathrm{g}(\beta))=\beta (f_N(\rho)-f_N(\rho_\mathrm{g} (\beta)))$.
Thus, minimizing the free-energy density with respect to the parameters $\{\phi_j\}$ yields a final state $\rho_\mathrm{f}(\tau)$ that is closest to $\rho_\mathrm{g}(\beta)$ in terms of the specific relative entropy.
In the thermodynamic limit, the following bound holds~\cite{mori2016macrostate}:
\begin{equation}
    s(\rho\|\rho_\mathrm{g}(\beta)) \coloneqq\lim_{N\to \infty} s_N(\rho\|\rho_\mathrm{g}(\beta)) \gtrsim (\Delta O)^2,
    \label{eq:sbound}
\end{equation}
where $\Delta O = \mathrm{Tr}[\hat{O}(\rho-\rho_\mathrm{g}(\beta))]$.
This relation implies that if $s(\rho\|\rho_\mathrm{g}(\beta))=0$, then $\rho$ and $\rho_\mathrm{g}(\beta)$ are indistinguishable in terms of local observables, even if $\rho \neq \rho_\mathrm{g}(\beta)$.

In the following, we investigate $s(\rho_\mathrm{f}(\tau)\|\rho_\mathrm{g}(\beta))$ and $s(\bar{\rho}(\tau,\tau_\mathrm{a}) \|\rho_\mathrm{g}(\beta))$ in nonintegrable and integrable spin chains.

\section{Results}\label{sec:result}
\subsection{Nonintegrable system}
We consider a spin chain model subject to a tilted magnetic field.
The Hamiltonian is given by
\begin{equation}
    \hat{H}_\mathrm{f}= g \sum_{i=1}^N \hat{\sigma}_{i}^z \hat{\sigma}_{i+1}^z + h_x \sum_{i=1}^N \hat{\sigma}_i^x + h_z \sum_{i=1}^N \hat{\sigma}_i^z.
\end{equation}
We set the parameters as $(g,h_x,h_z)=(1,0.9045,0.809)$, for which the ETH has been numerically verified~\cite{kim2014testing}, and impose a periodic boundary condition (i.e., $\hat{\sigma}_{N+1}^z=\hat{\sigma}_{1}^z$).
Based on the ETH argument, we choose a homogeneous initial state,
\begin{equation}
    \phi_1=\ldots=\phi_N= \phi.
    \label{eq:homogeneous}
\end{equation}

\subsubsection{Infinite operation time $\tau \to \infty$}\label{subsec:nonintegrable_infinite}
\begin{figure}[t]
    \centering
    (a)\\
    \includegraphics[width=0.9\linewidth]{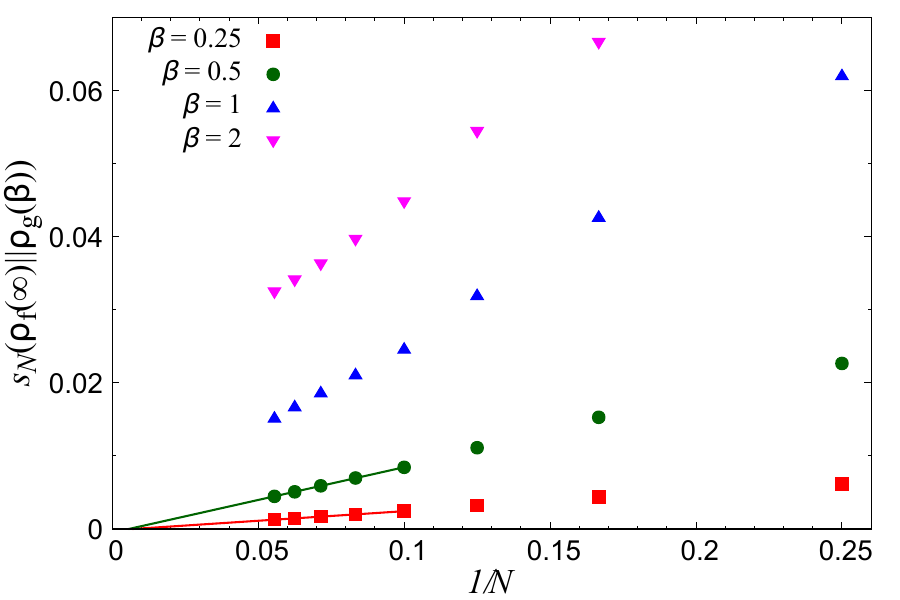}\\
    (b)\\
    \includegraphics[width=0.9\linewidth]{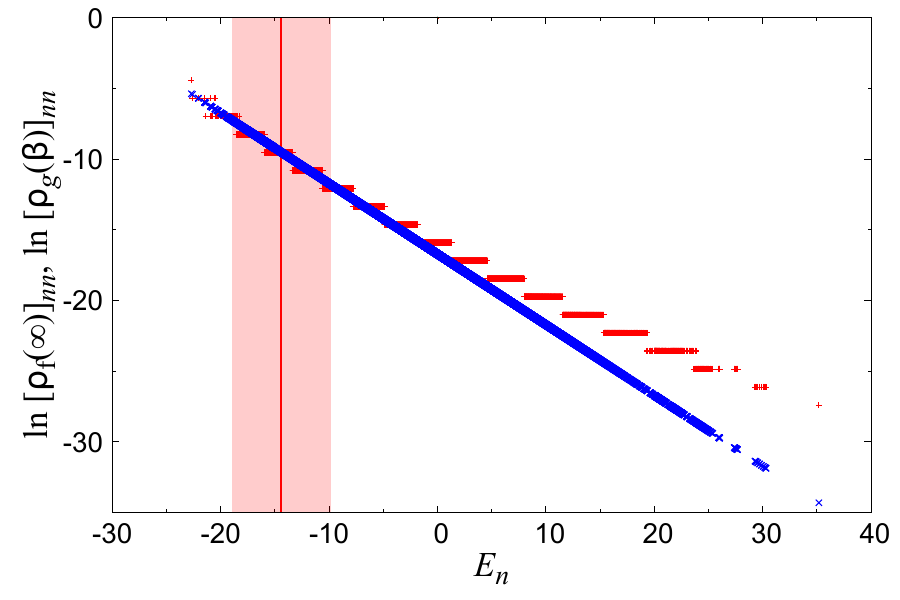}
    %(a)\\
    %\includegraphics[width=0.9\linewidth]{relative_beta4_tau_inf_v2.pdf}\\
    %(b)\\
    %\includegraphics[width=0.9\linewidth]{energy_density_diff_tau_inf_beta4.pdf}
    \caption{
    (a) System size dependence of $s_N(\rho_\mathrm{f}(\infty)\|\rho_\mathrm{g}(\beta))$.
    Squares (red), circles (green), up-triangles (blue), and down-triangles (magenta) denote the inverse temperature $0.25$, $0.5$, $1$, and $2$, respectively.
    %Both values exhibit the $N^{-1}$ scaling. 
    (b) Probability distribution of $\rho_\mathrm{f}(\infty)$ (red) and $\rho_\mathrm{g}(\beta)$ (blue) at $\beta=0.5$ and $N=18$.
    The vertical line and shaded region represent the energy expectation value $N e_N(\rho_\mathrm{f}(\infty))$ and the energy fluctuation $\delta E$, respectively.
    }
    \label{fig:tau_inf}
\end{figure}

We begin with the limit of infinite operation time $\tau \to \infty$.
In this adiabatic limit, $\rho_\mathrm{f}(\infty)$ is diagonal in the energy eigenbasis of $\hat{H}_\mathrm{f}$, with populations and their ordering preserved from the initial state within each subspace defined by translational symmetry~\footnote{We confirmed that the energy eigenvalues are nondegenerate within each subspace.}.
Figure~\ref{fig:tau_inf}~(a) shows the $N$ dependence of the specific relative entropy $s_N (\rho_\mathrm{f}(\infty)\|\rho_\mathrm{g}(\beta))$ for various values of $\beta$, where $\phi$ is determined by minimizing the free-energy density $f_N(\rho_\mathrm{f}(\infty))$.
%Figures~ display the probabilities $[\rho_\mathrm{f}(\infty)]_{nn}$ and $[\rho_\mathrm{g}(\beta)]_{nn}$ as a function of $E_n$.

At high temperatures (i.e., $\beta=0.25$ or $\beta=0.5$), we find linear dependence on $1/N$, so the value in the thermodynamic limit is estimated by extrapolating the data.
The extrapolated value almost vanishes, indicating the scaling:
\begin{equation}
    s_N (\rho_\mathrm{f}(\infty)\|\rho_\mathrm{g}(\beta)) \sim N^{-1}, N\gg 1.
    \label{eq:infinite_tau}
\end{equation}
Thus, in the limits $\tau \to \infty$ and $N\to \infty$, the process reproduces thermal properties.

To understand this scaling, we plot the optimized probability distribution of $\rho_\mathrm{f}(\infty)$ and the Gibbs state $\rho_\mathrm{g}(\beta)$ in Fig.~\ref{fig:tau_inf}~(b).
Now we label the energy eigenvalues $E_n$ by $E_m^{(\ell)}$ for $0\leq m\leq N$ and $1 \leq \ell \leq \tbinom{N}{m}$.
The energy eigenstates labeled by $m$ have the probability of $\cos^{2m}\phi \sin^{2(N-m)}\phi$ for $\rho_\mathrm{f}(\infty)$.

We define the energy window $W_m = [\min_\ell{E_m^{(\ell)}}, \max_\ell {E_m^{(\ell)}}]$ and assume that the overlap between $W_m$ and $W_{m+1}$ becomes small for large $N$ and $m\sim N$, both geometrically and in terms of state counting (see Appendix~\ref{appendix:energy_width} for quantitative analysis).
The ratio of the number of energy eigenvalues in $W_{m+1}$ to that in $W_m$ is given by
\begin{equation}
    \frac{\tbinom{N}{m+1}}{\tbinom{N}{m}}=\frac{N-m}{m+1},
\end{equation}
which is $O(1)$ for $m \sim N$.
Since the logarithm of the number of states gives the entropy, the entropy difference between $W_m$ and $W_{m+1}$ is finite.
Using the thermodynamic relation $\Delta S = \beta \Delta E$, we conclude that the width of each energy window $W_m$ is $O(1)$.
Hence, we can decompose the energy eigenvalues as $E_m^{(\ell)} = \bar{E}_m + \epsilon_m^{(\ell)}$, where
\begin{equation}
    \epsilon_m^{(\ell)} \sim O(1) \text{ for } m \sim N.
    \label{eq:width}
\end{equation}

For large $N$ at finite temperature, minimization of the free-energy density implies that the energy expectation value satisfies $Ne_N(\rho_\mathrm{f}(\infty)) - E_0 = O(N)$.
The energy fluctuation scales as $\delta E = O(\sqrt{N})$, as confirmed numerically (not shown).
Motivated by the behavior observed in Fig.~\ref{fig:tau_inf}~(b), we further assume that within the energy shell $|\bar{E}_m - Ne_N(\rho_\mathrm{f}(\infty))|\leq \delta E$, the quantity $\bar{E}_m$ can be locally approximated as
\begin{equation}
    \bar{E}_m=m\Delta+ N e_0,
    \label{eq:linear_app}
\end{equation}
where $\Delta$ and $e_0$ are constants.

Using Eqs.~(\ref{eq:width}) and~(\ref{eq:linear_app}), we can show that both free energy densities of $\rho_\mathrm{f}(\infty)$ and $\rho_\mathrm{g}(\beta)$ are given as
\begin{equation}
    f_N=e_0 - \beta^{-1} \ln (1+e^{-\beta\Delta}) +O(N^{-1}).
    \label{eq:free_energy_nonintegrable}
\end{equation}
Thus, the specific relative entropy becomes $O(N^{-1})$ for large $N$ and at finite temperature.

The agreement with the Gibbs state is consistent with the ETH.
%ETH is believed as a generic property of non-integrable models, stating that the expectation values of local observables are smooth functions of energy density.
Since the specific relative entropy is related to the free-energy density, Eq.~(\ref{eq:infinite_tau}) indicates that the difference in energy density vanishes as $N\to\infty$ as well as the difference in free-energy density.
Together with the ETH argument, this implies that $\rho_\mathrm{f}(\infty)$ is indistinguishable from the Gibbs state with respect to local observables as long as the energy fluctuation is macroscopically small.

At low temperatures (i.e., $\beta=1$ or $\beta=2$), we observe that the extrapolated value of the specific relative entropy remains finite.
We attribute this behavior to a finite-size effect.

The argument leading to the $N^{-1}$ scaling relies on the assumption that $Ne_N(\rho_\mathrm{f}(\infty))-E_0=O(N)$,
%However, in the low-temperature regime accessible to numerical simulations, the system is strongly localized around the ground state.
%, and the average energy remains $O(1)$.
%
which is guaranteed by free-energy minimization for large $N$ at finite temperature.
Indeed, for low-lying sectors with $m=O(1)$, the energy and entropy changes between neighboring energy windows $W_m$ and $W_{m+1}$ scale as $\Delta E=O(1)$ and $\Delta S = O(\ln N)$, respectively.
As the system size increases, the entropic contribution eventually compensates for the energy cost, favoring states with $m\sim N$.

For small system sizes and low temperatures, however, the system remains in a pre-asymptotic regime, where the entropic gain is insufficient to compensate for the energy cost, and the system is trapped in sectors with small $m$.
%This behavior reflects the fact that the system sizes  remain in a pre-asymptotic regime.
%, while free-energy minimization predicts $n=O(N)$ for large $N$.
We therefore expect a crossover from the low-temperature, ground-state-dominated regime to a high-temperature regime when $\beta \Delta E \lesssim \Delta S$.
Beyond this crossover, the average energy measured from $E_0$ becomes extensive, and the specific relative entropy exhibits the $N^{-1}$ scaling.
Using $\Delta S \sim \ln N$ for these low-lying sectors, the crossover system size scales as $N_\mathrm{c} \sim e^{\beta \Delta E}$.
For the low temperatures considered here, $N_\mathrm{c}$ is too large to be accessible to exact numerical simulations.

%At low temperatures (i.e., $\beta=1$ or $\beta=2$), the extrapolated value remains finite.
%The condition that the residual energy is $O(N)$ in the above argument does not hold, and $n\sim O(1)$.
%In this regime, the entropy change between $W_n$ and $W_{n+1}$ is $\ln N$.
%Thus, this regime is realized when
%$e \geq \beta^{-1} \ln N$,
%where $e$ is a typical energy change $E_{n+1}-E_n$, and $O(1)$.
%Thus, we expect that the low-temperature behavior crossovers to the high-temperature behavior when $N$ is sufficiently large, and the crossover occurs when $N\sim \exp (\beta e)$.

%The crossover between the high-temperature and low-temperature regimes occurs when the average energy crosses 
%Thus, the homogeneous initial states are insufficient to reproduce the thermal properties at low temperatures.
%It remains unclear whether this behavior is intrinsic or a finite-size effect artifact.

%, which exhibits the scaling:

%that the specific relative entropy decays inversely proportional to $N$ for large $N$.
%This implies that this process reproduces thermodynamic properties in the limits of $\tau \to \infty$ and $N\to \infty$.
%In the Appendix~\ref{appendix:linear}, we show the effect of adding one more parameter for representing the initial state and demonstrate its effectiveness to reduce the specific relative entropy for a given system size.

%In this subsection below, we set $\beta=0.5$

\subsubsection{Finite operation time}
\begin{figure}[t]
    \centering
    \includegraphics[width=0.9\linewidth]{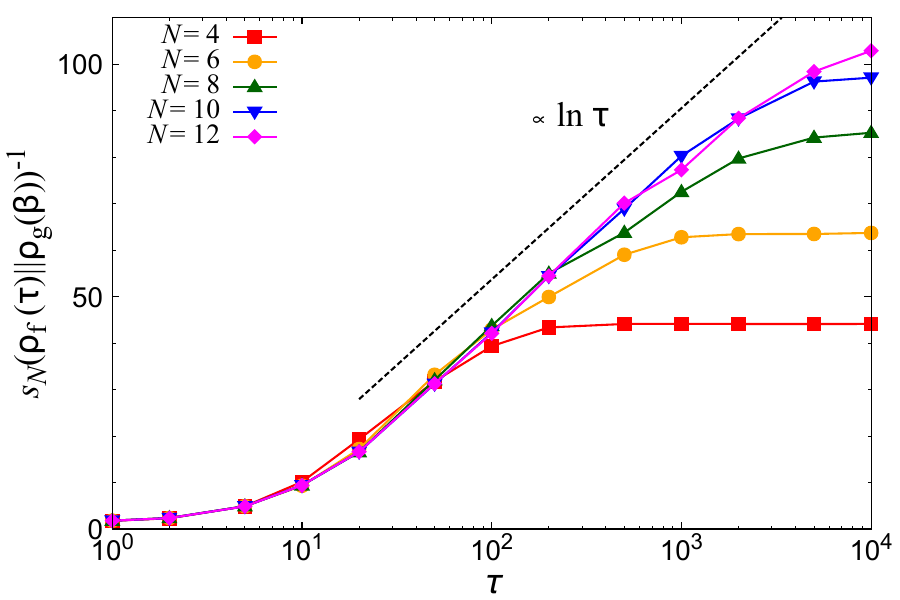}\\
    \caption{
    Dependence of the inverse of $s_N(\rho_\mathrm{f}(\tau)\|\rho_\mathrm{g}(\beta))$ on operation time $\tau$ for various system sizes.
    $\beta=0.5$ is chosen.
    The dotted line guides the scaling $s_N(\rho_\mathrm{f}(\tau)\|\rho_\mathrm{g}(\beta)) \sim (\ln\tau)^{-1}$.
    }
    \label{fig:tau_finite}
\end{figure}
Next, we investigate the operation-time dependence of $s_N(\rho_\mathrm{f}(\tau) \|\rho_{\rm g}(\beta))$, as shown in Fig.~\ref{fig:tau_finite}.
We adopt $\phi$ as the value optimized for each system size in the limit $\tau\to\infty$~\footnote{We adopt this choice to reduce the numerical costs.
Although the specific relative entropy becomes smaller when $\phi$ is optimized for each $\tau$, we have confirmed that the $\tau$-dependence qualitatively remains unchanged within the range of parameters we investigated.}.
We find that the specific relative entropy decreases with $\tau$ and eventually converges to the limiting value.
The saturation time $\tau_N$ increases with system size.
Before saturation, the quantity is independent of $N$, and the dependence of $\tau$ exhibits the scaling,
\begin{equation}
    s(\rho_\mathrm{f}(\tau)\|\rho_\mathrm{g}(\beta)) \sim (\ln\tau)^{-1}.
    \label{eq:finite_tau}
\end{equation}
%which indicates that reproducing thermal properties is not scalable.
This behavior can be understood as follows.
First, the saturated value obeys $s_N(\rho_\mathrm{f} (\tau_N) \|\rho_{\rm g} (\beta)) \sim N^{-1}$ (see Eq.~(\ref{eq:infinite_tau})).
Second, the saturation time $\tau_N$ is governed by the slowest timescale of the system set by the energy level spacings $\tau_N^{-1} \sim \exp(-\alpha N)$, where $\alpha > 0$.
Hence $\ln \tau_N \sim N$, and combining these relations yield Eq.~(\ref{eq:finite_tau}).

%Appendix~\ref{appendix:time_average} shows that a time-averaging process does not yield a state closer to the Gibbs state than $\rho_\mathrm{f}(\tau)$ in the thermodynamic limit.

\subsubsection{Time-averaging process}\label{sec:time_average}
\begin{figure}[t]
    \centering
%    (a)\\
%    \includegraphics[width=0.9\linewidth]{relative_temp4_finite_tau_inf_taum_linear.pdf}\\
%    (b)\\
    \includegraphics[width=0.9\linewidth]{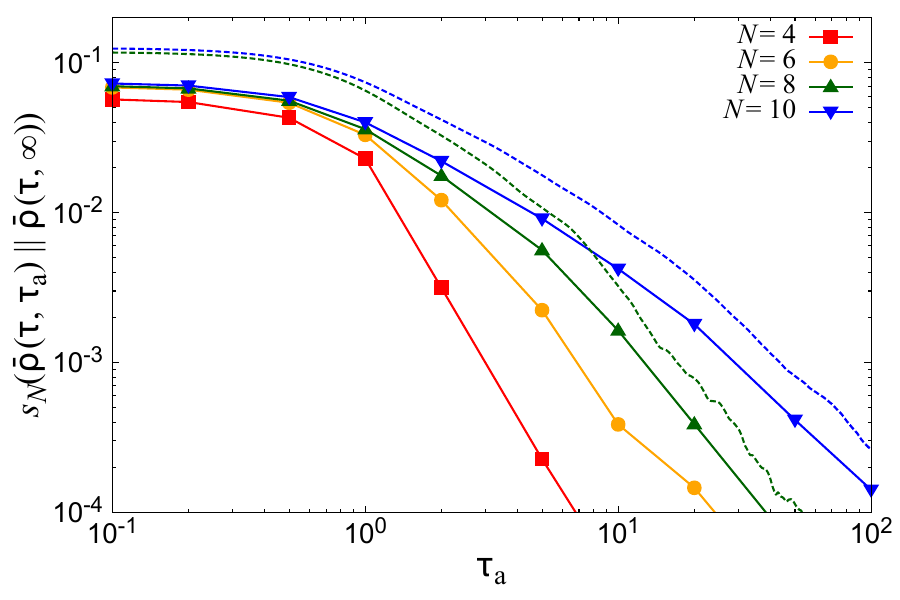}\\
    \caption{
%    (a) $\tau$-Dependence of $s_N(\bar{\rho}_*(\tau,\infty\|\rho_\mathrm{g}))^{-1}$ for various $N$.
    $\tau_\mathrm{a}$-Dependence of $s_N(\bar{\rho}(\tau,\tau_\mathrm{a}\| \bar{\rho}(\tau,\infty))$ at $\tau=10$ and $\beta=0.5$.
    The dashed lines provide bounds: $r_N(\bar{\rho}(\tau,\tau_\mathrm{a}\| \bar{\rho}(\tau,\infty))$.
    %The dotted and dashed lines are guides for the scaling $\tau_\mathrm{a}^{-1}$ and $\tau_\mathrm{a}^{-2}$, respectively.
    }
    \label{fig:taua}
\end{figure}

%Here, we consider a time-averaging process.
%Over an additional time window $\tau_\mathrm{a}$, we define the average density matrix
%\begin{equation}
%    \bar{\rho}(\tau,\tau_\mathrm{a}) = \int_0^{\tau_\mathrm{a}}  \rho_\mathrm{f} (\tau,t ) \frac{dt}{\tau_\mathrm{a}},
%\end{equation}
%where $\rho_\mathrm{f} (\tau,t ) = e^{-\mathrm{i}\hat{H}_\mathrm{f}t} \rho_\mathrm{f} (\tau ) e^{\mathrm{i}\hat{H}_\mathrm{f}t}$.
%This process reduces the free-energy density: $f_N(\bar{\rho}(\tau, \tau_\mathrm{a})) \leq f_N(\rho_\mathrm{f}(\tau))$.

One may expect that additional time averaging suppresses residual coherences in $\rho_\mathrm{f}(\tau)$ and thereby improves the approximation to the Gibbs state.
To examine this possibility, we consider the time-average density matrix $\bar{\rho}(\tau,\tau_\mathrm{a})$ in Eq.~(\ref{eq:time-average}).
In the limit of infinite averaging time $\tau_\mathrm{a} \to \infty$, all off-diagonal elements of $\rho_\mathrm{f} (\tau)$ vanish in the energy eigenbasis of $\hat{H}_\mathrm{f}$, and then
\begin{equation}
    \bar{\rho}(\tau, \infty)= \sum_n [\rho_\mathrm{f} (\tau)]_{nn} \ket{n}\bra{n}.
\end{equation}
The specific relative entropy can be decomposed as
\begin{align}
    &s_N(\bar{\rho}(\tau,\tau_\mathrm{a})\|\rho_\mathrm{g}(\beta))\nonumber\\
    =&s_N(\bar{\rho}(\tau,\infty)\|\rho_\mathrm{g}(\beta)) + s_N (\bar{\rho}(\tau,\tau_\mathrm{a}) \|\bar{\rho}(\tau,\infty)).
    \label{eq:decompose}
\end{align}
The first term is independent of $\tau_\mathrm{a}$.
The second term decreases with increasing $\tau_\mathrm{a}$ as shown in Fig.~\ref{fig:taua}, but the relaxation time to reach a fixed threshold value (e.g., $10^{-3}$) increases exponentially with system size.
This behavior is captured by an inequality that bounds the specific relative entropy from above by the R\'{e}nyi-2 entropy density~\cite{muller2013quantum} (i.e., $s_N(\rho\|\tau)\leq r_N(\rho\|\tau)$, where $r_N(\rho\|\tau) \coloneqq \ln [\mathrm{Tr} (\rho \tau^{-\frac{1}{2}}\rho \tau^{-\frac{1}{2}} )]/N$), which yields
%the time averaging suppresses the off-diagonal elements as
%\begin{equation}
%    [\bar{\rho}(\tau, \tau_\mathrm{a})]_{nm}=-\mathrm{i}\frac{1-e^{-2\mathrm{i} \Omega_{nm}}}{ 2\Omega_{nm}} [\rho_\mathrm{f}(\tau)]_{nm},
%\end{equation}, yield
\begin{equation}
    e^{N s_N (\bar{\rho}(\tau,\tau_\mathrm{a}) \|\bar{\rho}(\tau,\infty))} \leq 1+ \sum_{\substack{n,m\\ (n\neq m)}} g(\omega_{nm} \tau_\mathrm{a}) h_{nm} ,
\end{equation}
where $g(x)=(\sin x/x)^2$, $\omega_{nm}=(E_n-E_m)/2$ and $h_{nm}=|[\rho_\mathrm{f}(\tau)]_{nm}|^2/([\rho_\mathrm{f}(\tau)]_{nn}[\rho_\mathrm{f}(\tau)]_{mm})^{1/2}$.
As shown in Fig.~\ref{fig:taua} by dashed lines, the R\'{e}nyi-2 entropy density qualitatively reproduces the specific relative entropy.
The nonzero specific relative entropy at $\tau_\mathrm{a}=0$ implies that $\sum_{n,m(n\neq m)}h_{nm}$ is exponentially large with respect to $N$.
The correction at finite $\tau_\mathrm{a}$ is controlled by the oscillatory suppression of $g(\omega_{nm}\tau_\mathrm{a})$.
To quantitatively study the $\tau_\mathrm{a}$ dependence at large $N$, we introduce $D(\omega)=\sum_{n,m (n\neq m)} h_{nm} \delta(\omega - \omega_{nm})$ and assume that $D(\omega)=D$ for $\omega \in [-\Lambda,\Lambda]$ and otherwise $0$ with $\Lambda$ being a cutoff frequency.
Then
\begin{equation}
    \sum_{\substack{n,m\\ (n\neq m)}} g(\omega_{nm} \tau_\mathrm{a}) h_{nm} = D \tau_\mathrm{a}^{-1} \int_{-\Lambda \tau_\mathrm{a}}^{\Lambda \tau_\mathrm{a}} g(x) dx \sim \tau_\mathrm{a}^{-1}.
\end{equation}
%Since $g(x)\leq 1$ when $x\leq 1$ and $g(x)\leq x^{-2}$ when $x\geq 1$, 
These arguments indicate a transient timescale, which is exponentially long with respect to $N$, and the specific relative entropy behaves as
\begin{equation}
s_N(\rho_\mathrm{f}(\tau)\|\bar{\rho}(\tau,\infty))-s_N(\bar{\rho}(\tau,\tau_\mathrm{a})\|\bar{\rho}(\tau,\infty))\sim \frac{\ln \tau_\mathrm{a}}{N}.
\end{equation}

These results indicate that although time averaging gradually suppresses coherences, the associated relaxation timescale grows exponentially with system size.
Consequently, in the thermodynamic limit, time averaging does not reduce the specific relative entropy with respect to the Gibbs state beyond that achieved
by $\rho_\mathrm{f}(\tau)$ itself.

\subsection{Integrable system}
We next analyze the transverse-field Ising model~\cite{sachdev1999quantum}, which serves as a representative example of an integrable model:
\begin{equation}
    \hat{H}_\mathrm{f}=-J\sum_{i=1}^{N-1} \hat{\sigma}_i^z \hat{\sigma}_{i+1}^z - \Gamma \sum_{i=1}^N \hat{\sigma}_i^x,
\end{equation}
with an open boundary condition.
The free-energy density of the Gibbs state in the thermodynamic limit is given by~\cite{lieb1961two}
\begin{equation}
    f(\rho_\mathrm{g}(\beta))= - \frac{1}{\pi\beta} \int_0^\pi \ln[2\cosh (\beta \Lambda(y))] dy,
    \label{eq:free_energy_integrable}
\end{equation}
where $f(\rho)=\lim_{N\to \infty}f_N(\rho)$ and $\Lambda(y)=\sqrt{\Gamma^2+J^2+2\Gamma J \cos y}$.
The Hamiltonian used in the quasi-adiabatic thermal process is also integrable and can be written as
\begin{equation}
    \hat{H}(s_\mathrm{a})=\sum_{k=1}^N \Lambda_k(s_\mathrm{a}) \left( \hat{\eta}^\dagger_k (s_\mathrm{a}) \hat{\eta}_k (s_\mathrm{a}) - \frac{1}{2} \right),
\end{equation}
where $\hat{\eta}_k(s_\mathrm{a})$ and $\hat{\eta}_k^\dagger(s_\mathrm{a})$ are fermionic annihilation and creation operators, and $\{\Lambda_k(s_\mathrm{a})\}_{k=1}^N$ denotes the single particle spectrum, ordered in ascending order.

\subsubsection{Homogeneous initial state}
First, we consider $\rho_\mathrm{f}(\tau)$ starting from a homogeneous initial state (i.e.,  $\phi_1=\ldots=\phi_N=\phi$).
The parameter $\phi$ is determined to minimize the free-energy density $f_N(\rho_\mathrm{f}(\tau))$ at a given $\beta$ and $N$.
Under the Jordan--Wigner transformation~\cite{lieb1961two}, the Hamiltonian is written as
\begin{equation}
    \hat{H}_\mathrm{f}=-J\sum_{i=1}^{N-1} (\hat{c}_i^\dagger -\hat{c}_i)(\hat{c}_{i+1}^\dagger +\hat{c}_{i+1})-\Gamma \sum_{i=1}^N (2 \hat{c}_i^\dagger \hat{c}_i -1),
\end{equation}
where $\hat{c}_i$ and $\hat{c}_i^\dagger$ are annihilation and creation fermion operators.
The energy density of $\rho_\mathrm{f}(\tau)$ can be expressed as
\begin{align}
    e(\rho_\mathrm{f}(\tau))
    = \sum_{j=1}^N \sum_{k=1}^{N} \frac{\Lambda_j(1)}{2N} (|u_{j,k}|^2-|v_{j,k}|^2) \cos 2\phi,
    \label{eq:energy_integrable}
\end{align}
where $\hat{U}_\tau^\dagger \hat{\eta}_i(1) U_\tau= \sum_{j=1}^N [u_{i,j} \hat{c}_j + v_{i,j}^* \hat{c}_j^\dagger]$~\cite{caneva2007adiabatic}.
When $\phi=0$, the energy density is identical to that of quantum annealing, denoted by $e_\mathrm{a}(\tau)$.
Therefore, we obtain
\begin{equation}
    e(\rho_\mathrm{f}(\tau))=e_\mathrm{a}(\tau) \cos 2\phi.
    \label{eq:relation_QA}
\end{equation}
Then, minimizing the free-energy density gives
\begin{equation}
    f_N(\rho_\mathrm{f}(\tau))=-\frac{1}{\beta} \ln [2 \cosh( \beta e_\mathrm{a}(\tau))].
\end{equation}

For the present model, in the thermodynamic limit, quantum annealing reproduces the ground-state energy density in the limit of infinite operation time~\cite{dziarmaga2005dynamics}.
%In the infinite operation time, quantum annealing reproduces the ground state~\cite{morita2008mathematical}: $\lim_{\tau \to \infty} e_\mathrm{a}(\tau)=-\sum_k \Lambda_k(1)/2$.
Thus, the free-energy density becomes
\begin{equation}
    f(\rho_\mathrm{f} (\infty)) = - \frac{1}{\beta} \ln \left[2\cosh \left(\frac{\beta}{\pi} \int_0^\pi \Lambda(y)dy\right) \right].
    \label{eq:free_energy_homogeneoud_integrable}
\end{equation}
Compared to Eq.~(\ref{eq:free_energy_integrable}), the homogeneous initial state does not reproduce the thermal properties (i.e., $s(\rho_\mathrm{f} (\infty)\| \rho_\mathrm{g}(\beta)) \neq 0$) except when $\Gamma J = 0$ or $\beta=0$.

When $\Gamma=0$ at finite operation time, the energy density of quantum annealing is given by~\cite{dziarmaga2005dynamics}
\begin{equation}
    e_\mathrm{a}(\tau)= -J\left(1-\frac{2}{\pi\sqrt{J\tau}} \right).
\end{equation}
The $\tau$ dependence is explained by the Kibble-Zurek mechanism~\cite{zurek2005dynamics} and also appears in the specific relative entropy as
\begin{equation}
    s(\rho_\mathrm{f}(\tau) \| \rho_\mathrm{g}(\beta)) \simeq \frac{2 \beta}{\pi} \sqrt{\frac{J}{\tau}} \tanh{(\beta J)},
    \label{eq:relative_entropy_integrable}
\end{equation}
which is derived by the Taylor expansion of the free-energy density with respect to $J$.

To understand the results, we interpret them in terms of $W_m$ and $\bar{E}_m$, introduced in~\ref{subsec:nonintegrable_infinite}.
There, we show that the specific relative entropy vanishes in the limit of $\tau\to \infty$, if (i) the width of $W_m$ is finite and (ii) $\bar{E}_m$ is locally given by Eq.~(\ref{eq:linear_app}).
When $\Gamma=0$, the single-particle energies $\{\Lambda_k(1)\}$ are degenerate except for a mode representing the spontaneously symmetry breaking (i.e., $\Lambda_1(1)=0$ and $\Lambda_k(1)=2J$ for $k\geq 2$).
Thus, the first condition is satisfied: the widths of $W_m$ are $0$ for $m=0$ and $2J$ for $1\leq m \leq N$.
The second condition is satisfied regardless of $m$ when taking $\Delta=2J$ and $e_0=-J$.
Then, Eq.~(\ref{eq:free_energy_nonintegrable}) reproduces the exact free-energy density in Eq.~(\ref{eq:free_energy_integrable}) in the thermodynamic limit.
Therefore, the specific relative entropy vanishes in the limit of $\tau\to \infty$ despite the homogeneous initial state.
However, for generic $\Gamma \neq 0$, the single-particle energies are non-degenerate, leading to an extensive width of $W_m$ for $m\sim N$.
As a result, the specific relative entropy generally remains finite even in the limit $\tau \to \infty$ when using a homogeneous initial state.

\subsubsection{Inhomogeneous initial state}
%\begin{figure}[t]
%    \centering
    %(a)\\
    %\includegraphics[width=0.9\linewidth]{single_critical.pdf}\\
    %(b)\\
%    \includegraphics[width=0.9\linewidth]{single.pdf}
%    \caption{
%    s
%    }
%    \label{fig:integrable}
%\end{figure}
\begin{figure}[t]
    \centering
    (a)\\
    \includegraphics[width=0.9\linewidth]{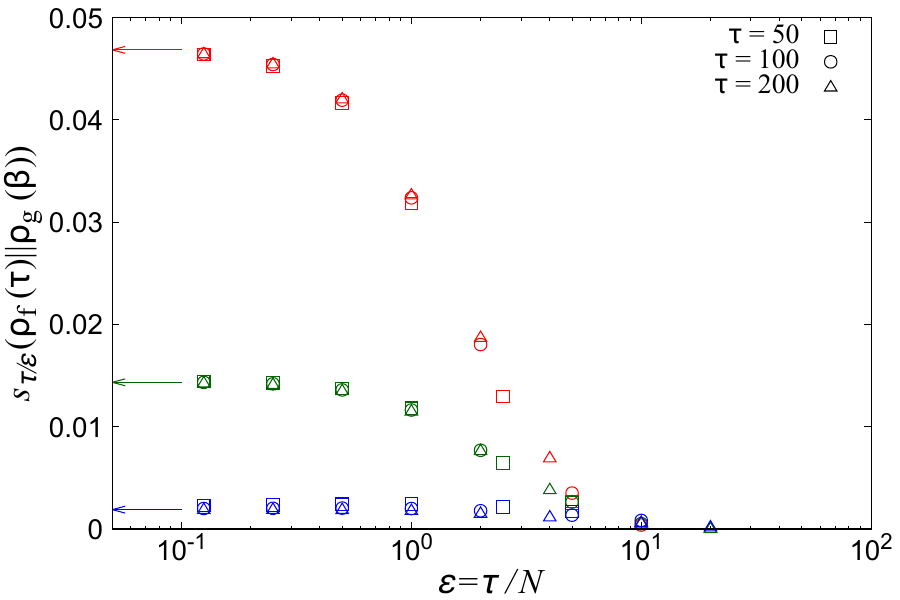}\\
    (b)\\
    \includegraphics[width=0.9\linewidth]{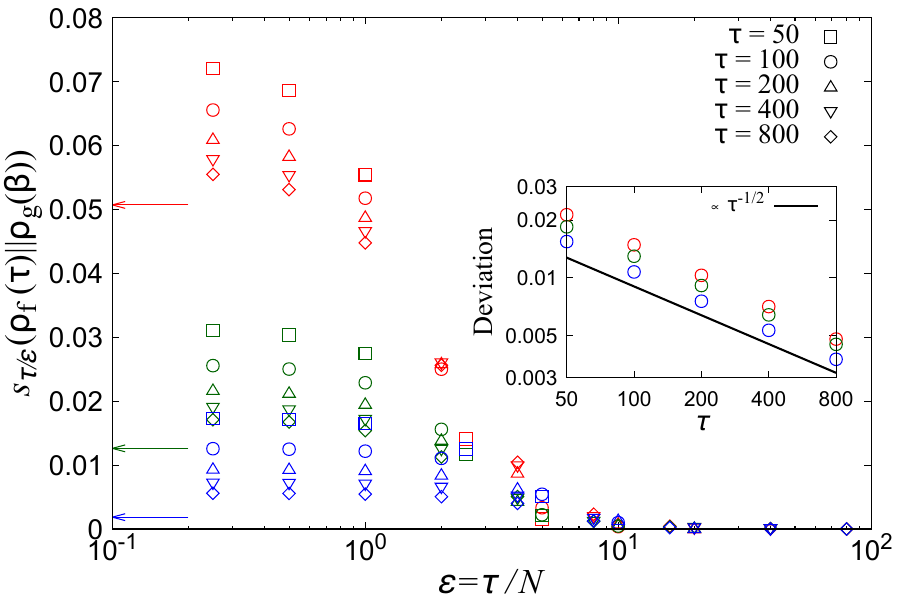}
    \caption{
    Dependence of $s_{\tau/\epsilon}(\rho_\mathrm{f}(\tau)\|\rho_\mathrm{g}(\beta))$ on $\epsilon$ for various values of $\tau$ and $n_p$ in (a) paramagnetic phase $(J, \Gamma)=(1, 2)$ and (b) ferromagnetic phase $(J, \Gamma)=(1, 0.5)$.
    The cases $n_p=1,2$ and $5$ are indicated by red, green, and blue symbols, respectively, while different symbol shapes correspond to different values of $\tau$.
    $\beta = 1$ is chosen.
    The arrows indicate the lower bounds $s_\mathrm{b}$, given in Eq.~(\ref{eq:bound}).
    (inset) $\tau$ dependence of the deviation, $s_{\tau/\epsilon}(\rho_\mathrm{f}(\tau)\|\rho_\mathrm{g}(\beta))-s_\mathrm{b}$, at $\epsilon=0.25$.
    }
    \label{fig:integrable}
\end{figure}

We now consider the case of inhomogeneous initial state.
%In this subsection, we assume that $\phi_1 > \phi_2 > \ldots > \phi_N$ and $\{\Lambda_k(s_\mathrm{a})\}_{k=1}^N$ are nondegenerate for $s_\mathrm{a}\in[0,1]$.
In this subsection, we assume that $\{\Lambda_k(s_\mathrm{a})\}_{k=1}^N$ is nondegenerate for $s_\mathrm{a}\in[0,1]$.
The energy density of $\rho_\mathrm{f}(\tau)$ is given by
\begin{equation}
    e_N(\rho_\mathrm{f}(\tau))=\frac{1}{N} \sum_{k=1}^N \Lambda_k(1) \left[\mathrm{Tr} (\hat{\eta}_k^\dagger(1)\hat{\eta}_k(1) \rho_\mathrm{f}(\tau))-\frac{1}{2}\right].
\end{equation}
For finite $N$ and in the infinite limit of $\tau$, the adiabatic theorem implies that, for $k\in \{1,\ldots,N\}$,
\begin{equation}
    \hat{\eta}_k^\dagger(1)\hat{\eta}_k(1) \rho_\mathrm{f}(\infty) =\hat{\eta}_k^\dagger(0)\hat{\eta}_k(0)\rho_\mathrm{i},
\end{equation}
and then
\begin{equation}
    e_N(\rho_\mathrm{f}(\infty))= -\frac{1}{2N} \sum_{k=1}^N \Lambda_k(1) \cos (2\phi_{\sigma(k)}),
\end{equation}
where $\sigma(k)$ is a permutation and $\phi_{\sigma(1)} > \phi_{\sigma(2)} > \ldots > \phi_{\sigma(N)}$.
The free-energy density coincides with that of the Gibbs state when 
\begin{equation}
    \phi_{\sigma(k)} = \frac{1}{2} {\rm arccos} \left[ \tanh \left( \frac{\beta \Lambda_k(1)}{2} \right) \right],
    \label{eq:condition}
\end{equation}
which corresponds to the condition that $\hat{H}_\mathrm{i}$ and $\hat{H}_\mathrm{f}$ share the same single-particle spectra up to the constant factor~\cite{irmejs2025quasi,plastina2014irreversible}:
\begin{equation}
    \beta_\mathrm{i} \Lambda_k (0) =\beta \Lambda_k(1).
\end{equation}
%This is called  condition in
This observation implies that, unlike nonintegrable case, an extensive number of parameters is generally required to reproduce the thermal properties of local observables in the integrable case.

%Finally, we discuss $\tau$ dependence of the specific relative entropy at $\Gamma=0$.
%In the limit of $\tau \to \infty$, 
%\begin{equation}
%    f(\rho_*(\infty))=f(\rho_\mathrm{g})=-\frac{1}{\beta} \ln (2\cosh \beta J),
%\end{equation}
%and $\phi_i = \arccos{(\frac{1}{2} \tanh{(\beta J)})}$ is homogeneous.
%When $\phi_i$ is assumed to be homogeneous, the free energy density is given as (see Appendix~\ref{appendixA})
%\begin{equation}
%    f(\rho(\tau))=\cos (2\phi) e_\mathrm{a} (\tau) -\beta^{-1} s(\phi),
%\end{equation}
%where $e_\mathrm{a}(\tau)$ denotes the energy density in quantum annealing~\cite{dziarmaga2005dynamics}:

%Therefore,
%\begin{align}
%    f(\rho_*(\tau))\leq & -\frac{1}{\beta} \ln \left(2\cosh \left[\beta J\left(1-\frac{2}{\pi\sqrt{J\tau}} \right) \right] \right) \nonumber\\
%    \simeq & f(\rho_g) + \frac{2}{\pi} \sqrt{\frac{J}{\tau}} \tanh{(\beta J)} \nonumber\\
%    \Leftrightarrow s(\rho_*(\tau) \| \rho_\mathrm{g}) \lesssim &\frac{2 \beta}{\pi} \sqrt{\frac{J}{\tau}} \tanh{(\beta J)}
%\end{align}

Next, we investigate the $\tau$ dependence by setting $\{\phi_k\}$ according to Eq.~(\ref{eq:condition}).
We select the permutation as
\begin{equation}
    \sigma (\lceil N/n_p\rceil (i-1)+j)=n_p(j-1)+i.
\end{equation}
where $1 \leq i \leq n_p$ and $1 \leq j \leq \lceil N/n_p\rceil$.
The permutation generates a pseudo-periodic structure of transverse fields with spatial period $n_p$,
which becomes locally periodic in the thermodynamic limit.
The parameter $n_p$ therefore controls the degree of spatial modulation in the initial state:
$n_p=1$ corresponds to a monotonic structure that varies smoothly.
In the thermodynamic limit, the variation occurs on a macroscopic scale, so that the state is locally uniform while remaining globally inhomogeneous.
Larger $n_p$ introduce finer spatial modulation in the local structure.
This construction allows us to examine how spatial structure in the initial state affects the approach to thermal equilibrium.

%The energy density of $\rho_\mathrm{f}(\tau)$ is obtained by the replacement of $\cos \phi$ in Eq.~(\ref{eq:energy_integrable}) by $\cos \phi_k$.
%Taking the $\tau \to \infty$ limit with fixed $\epsilon$ under the assumption of finite $n_p$~\footnote{When $n_p$ diverges with $N$, the specific relative entropy reduces to the result of $n_p=1$.},
In the thermodynamic limit, we obtain a lower bound on the specific relative entropy (see Appendix~\ref{appendix:integrable} for the derivation): for finite $n_p$
%\begin{align}
%    &\lim_{\tau\to\infty}s_{\tau/\epsilon}(\rho_\mathrm{f}(\tau)\|\rho_\mathrm{g}(\beta)) \nonumber\\
%    = & \beta \int_0^\pi \frac{dy}{\pi} \left(\Lambda(y)-\int_0^\pi \frac{dx}{\pi} \Lambda(x)\right) \tanh (\beta \Lambda(y))+O(\epsilon),
%    \label{eq:bound}
%\end{align}
\begin{align}
    &\lim_{\tau\to\infty} s(\rho_\mathrm{f}(\tau)\|\rho_\mathrm{g}(\beta)) \nonumber\\
    \geq & \beta \sum_{n=1}^{n_p} \int_{\frac{n-1}{n_p}\pi}^{\frac{n}{n_p}\pi} \frac{dy}{\pi} \tanh (\beta \Lambda(y)) \left( \Lambda(y)- \int_{\frac{n-1}{n_p}\pi}^{\frac{n}{n_p}\pi}  \frac{\Lambda(x) dx}{\pi/n_p} \right) \nonumber\\
    \eqqcolon & s_\mathrm{b}.
    %& \times \left( \Lambda(y)- \int_{\frac{n-1}{n_p}\pi}^{\frac{n}{n_p}\pi}  \frac{\Lambda(x) dx}{\pi/n_p} \right),
    \label{eq:bound}
\end{align}
%which follows from the locality and the translational symmetry of $\hat{H}_\mathrm{f}$.
We make two remarks.
First, the inequality becomes an equality when $n_p=1$.
Second, the limits $N\to\infty$ and $\tau\to\infty$ do not commute (i.e., taking $\tau\to\infty$ first and then $N\to\infty$ yields a vanishing specific relative entropy).
The non-commutativity originates from the finite propagation velocity implied by the Lieb--Robinson bound, reflecting the locality of $\hat{H}(s_\mathrm{a})$.
%Note that we assume $n_p =O(N^0)$
%Note that this bound is different from $\lim_{N\to\infty}s_N(\rho_\mathrm{f}(0)\|\rho_\mathrm{g}(\beta))$ (i.e., the two limits $N\to \infty$ and $\epsilon \to 0$ do not commute).

Figures~\ref{fig:integrable}~(a) and~(b) show the specific relative entropy $s_{\tau/\epsilon}(\rho_\mathrm{f}(\tau)\|\rho_\mathrm{g}(\beta))$ as a function of $\epsilon=\tau/N$ for various values of $\tau$ and $n_p$.
The thermodynamic limit corresponds to the value in the limit of $\epsilon \to 0$ and the bounds in Eq.~(\ref{eq:bound}) are indicated by arrows.
In the paramagnetic phase $(J,\Gamma)=(1,2)$, as shown in Fig.~\ref{fig:integrable}~(a), at small $\epsilon$, the specific relative entropy converges to the bound.
In contrast, in the ferromagnetic phase $(J,\Gamma)=(1,0.5)$, as shown in Fig.~\ref{fig:integrable}~(b), a visible deviation from the bound remains even at $\tau=800$.
The deviation decays as $\tau^{-1/2}$~\footnote{The fitting gives exponents of $-0.54, -0.51$, and $-0.51$ for $n_p=1, 2$, and $5$, respectively.
The small deviation from $-1/2$ is attributed to finite $\epsilon$.}, which is explained by the Kibble--Zurek mechanism (see Appendix.~\ref{appendix:integrable} and~\ref{appendix:universality}).
Thus, we numerically found that the bound in Eq.~(\ref{eq:bound}) is tight in the limit of $\tau \to \infty$ for all the values of $n_p$ considered.

Thus, we conclude that the specific relative entropy can be decomposed into two contributions:
\begin{equation}
    s(\rho_\mathrm{f}(\tau)\|\rho_\mathrm{g}(\beta))=s_\mathrm{b} + \Delta s(\tau).
\end{equation}
The Taylor expansion yields $s_\mathrm{b} \sim n_p^{-2}$.
Physically, increasing $n_p$ suppresses the deviation arising from the globally inhomogeneous initial state.
$\Delta s(\tau)$ vanishes as $\tau \to \infty$ and its decay is governed by the Kibble--Zurek mechanism if a quantum phase transition exists during the quasi-adiabatic thermal process.
Thus, both large $n_p$ and long $\tau$ are required to generate thermal ensembles close to the Gibbs state.
%It should be noted that, unlike non-integrable systems, the performance of the quasi-adiabatic thermal process is affected by the presence of a quantum phase transition.

%and ferromagnetic phase $(J,\Gamma)=(1,0.5)$.
%As shown in Fig.~\ref{fig:integrable}~(a), in the paramagnetic phase, at small $\epsilon$, the specific relative entropy converges to the asymptotic value, $\lim_{\epsilon \to 0}(\lim_{\tau\to\infty}s_{\tau/\epsilon}(\rho_\mathrm{f}(\tau)\|\rho_\mathrm{g}(\beta)))$, indicated by an arrow, at relatively small $\tau\simeq 20$.
%In contrast, a visible deviation remains even at $\tau=800$ in the ferromagnetic phase.
%As shown in the inset of Fig.~\ref{fig:integrable}~(b), the deviation decays as $\tau^{-1/2}$~\footnote{The fitting gives an exponent of $-0.52$, and the deviation from $-1/2$ is attributed to finite $\epsilon$.}, which is explained by the Kibble--Zurek mechanism (see Appendix.~\ref{appendix:integrable}).
%For large $\tau$, the specific relative entropy approaches a function of $\epsilon$ and vanishes at large $\epsilon$ in both figures, indicating that the operation time scales linearly with $N$ (i.e., diverges in the thermodynamic limit) to achieve a small precision threshold.
%Physically, it represents a time that a local operator spreads over the entire system.

\section{Conclusion}\label{sec:conclusion}
We investigate the quasi-adiabatic thermal process for preparing thermal ensembles in the thermodynamic limit.
For nonintegrable spin chains, our numerical results combined with a thermodynamic argument indicate that a single parameter is sufficient to reproduce the thermal properties of local observables, although the operation time increases exponentially with precision.
We also find that time averaging does not improve performance.
%We numerically demonstrate its effectiveness for the nonintegrable spin chains.
In the integrable transverse-field Ising model, we clarify the differences from the nonintegrable case, which arise from the presence of local conserved quantities.
In general, fine-tuning of the initial state is required, and the performance is affected by the presence of a quantum phase transition.

From a broader perspective, thermal ensemble preparation has been studied using both classical and quantum approaches.
%Classical methods, such as quantum Monte Carlo, tensor-network techniques, and machine-learning approaches, are powerful but often suffer from fundamental limitations, including the sign problem, entanglement growth, or restricted expressibility.
%Quantum approaches, including the quantum Metropolis algorithm, imaginary-time evolution, Lindblad dynamics simulation, and variational methods, offer alternative routes but typically require deep circuits, engineered dissipation, or carefully designed ansatz.
Unlike classical methods, the quasi-adiabatic process encodes thermal properties through coherent quantum dynamics.
Compared with existing quantum algorithms for thermal state preparation, our approach is purely unitary and does not rely on phase estimation, engineered dissipation, or explicit entropy estimation.
This makes it potentially suitable for near-term quantum implementations.
At the same time, our analysis reveals intrinsic limitations governed by adiabatic timescales and the presence of integrability.
%In this context, the quasi-adiabatic thermal process studied here provides a complementary framework: it avoids explicit imaginary-time evolution and dissipative engineering while naturally incorporating quantum correlations, although its efficiency is constrained by adiabatic timescales and critical phenomena.

These results highlight both the potential and limitations of quantum approaches based on quasi-adiabatic thermal process for thermal state preparation.
An open question is how these results extend to other  systems, such as integrable models solvable by the Bethe ansatz or systems in higher spatial dimensions, where phase transitions occur at finite temperature.

%An open question concerns the limitations of the proposed algorithm in the low-temperature regime.
%Our results indicate that a time averaging process is necessary to reproduce the thermal properties with a polynomial cost, but this process inevitably increases the entropy density.
%This implies that the algorithm has an intrinsic limitation when applied to the low-temperature regime, which is required for further study. 
%Extending this approach to systems with higher spatial dimensions is an intriguing direction of research.
%In addition, our results motivate the development of a hybrid architecture that combines gate-based and annealing quantum processors.

\begin{acknowledgments}
This work was supported by JSPS KAKENHI (Grant Number 23K13034).
The numerical calculations were partly supported by the supercomputer center of ISSP of Tokyo University.
\end{acknowledgments}

%{\it Note added --} Recently, we became aware of related work~\cite{granet2025adiabatic}.
\appendix
%\begin{appendices}
\twocolumngrid

\section{Overlap of the energy window}\label{appendix:energy_width}
\begin{table}[t]
  \centering 
  \caption{Overlap of energy window.
   } 
  \scalebox{1}{
  \begin{tabular}{cccc|cccc}\hline
   $N$ & $m_*$ & $r_\mathrm{c}$ & $r_\mathrm{w}$ & $N$ & $n_*$ & $r_\mathrm{c}$ & $r_\mathrm{w}$ \\ \hline\hline
   4 & 1 & 0.0 & 0.0 & 12 & 3 & 0.045 & 0.087\\ 
   6 & 1 & 0.0 & 0.0 & 14 & 3 & 0.049 & 0.129\\
   8 & 2 & 0.036 & 0.082 & 16 & 4 & 0.026 & 0.026\\
   10 & 2 & 0.089 & 0.058 & 18 & 4 & 0.012 & 0.020\\
   \hline
   \end{tabular}
  }
  \label{Table:overlap}
\end{table}

This appendix numerically verifies the assumption made in the main text that the overlap of the energy windows $W_m = [\min_\ell E_m^{(\ell)},\max_\ell E_m^{(\ell)}]$ is small in the nonintegrable model at $\beta=0.5$.
We first identify $W_{m_*}$ whose center is the closest to the average energy of the final state,
\begin{equation}
    m_*=\mathrm{argmin}_m \left|N e_N(\rho_\mathrm{f}(\infty))-\frac{\min_\ell E_m^{(\ell)}+\max_\ell E_m^{(\ell)}}{2} \right|.
\end{equation}
We then quantify the overlap between $W_{m_*}$ and the others using two complementary measures: a state-counting overlap and a geometric overlap.
The state-counting overlap is defined as
\begin{equation}
    r_\mathrm{c}=\frac{\# \{ E_{m}^{(\ell)} \in W_{m_*} | m\neq m_*\}}{\# \{E_{m_*}^{(\ell)} \}},
\end{equation}
which measures the fraction of energy eigenvalues within $W_{m_*}$ that originate from the different $m$-sectors.
The geometric overlap is defined as
\begin{equation}
    r_\mathrm{w}=\frac{d \left(W_{m_*} \cap \left(\bigcup_{\substack{m=0\\(m\neq m_*)}}^N W_m\right) \right)}{d(W_{m_*})},
\end{equation}
where $d(A)$ denotes the Lebesgue measure (total length) of a set $A \subset \mathbb{R}$.
In particular, for a single interval $A=[a,b]$, we have $d(A)=|b-a|$.
The assumption used in the main text is validated if both $r_\mathrm{c}$ and $r_\mathrm{w}$ become smaller than unity for large $N$.

Table~\ref{Table:overlap} shows that both $r_\mathrm{c}$ and $r_\mathrm{w}$ are significantly smaller than unity for all $N$ accessible in our numerical calculations.
This behavior supports the assumption that the overlap between energy windows is small in the thermodynamic limit.
In addition, we find $m_* \sim N$, consistent with the scaling $N e_N(\rho_\mathrm{f}(\infty))-E_0=O(N)$.

\section{Derivation of Eq.~(\ref{eq:bound})}\label{appendix:integrable}
We derive Eq.~(\ref{eq:bound}) by extending the zero-temperature result of~\cite{mbeng2019quantum} to finite temperatures.
Since the specific relative entropy can be expressed in terms of the free-energy density difference, and the entropy contribution vanishes when $\{\phi_j\}$ is set according to Eq.~(\ref{eq:condition}), it suffices to evaluate the local energy associated with the sites between $(j-1)n_p+1$ and~$jn_p$:
\begin{equation}
    e_{j}=\mathrm{Tr} [\hat{U}_\tau^{\dagger} \hat{h}_{j}(J,\{\Gamma\}) \hat{U}_\tau \rho_\mathrm{i}],
    \label{eq:energyj}
\end{equation}
%\begin{equation}
%    e(\rho_\mathrm{f}(\tau))=\frac{1}{N} \sum_{j=1}^{N-1} \mathrm{Tr} [\hat{U}_\tau^{\dagger} \hat{h}_j \hat{U}_\tau \rho_\mathrm{i}]+\frac{\Gamma}{N}\mathrm{Tr} [\hat{U}_\tau^{\dagger} \hat{\sigma}_N^z \hat{U}_\tau \rho_\mathrm{i}],
%\end{equation}
where $\hat{h}_{j} (J,\{\Gamma_k\})=\sum_{k=1}^{n_p}(-J \hat{\sigma}_{(j,k)}^z \hat{\sigma}_{(j,k+1)}^z-\Gamma_k \hat{\sigma}_{(j,k)}^x)$ and $(j,k)=(j-1) n_p+k$.
%It is noted that the difference between $e_\mathrm{p}$

%First we introduce a Hamiltonian
%with a periodic boundary condition:
%Then the energy density is given as
%\begin{equation}
%    e(\rho_\mathrm{f}(\tau))=\frac{1}{N} \sum_{j=1}^{N} \mathrm{Tr} [\hat{U}_\tau^{\dagger} \hat{h}_j(J,\Gamma) \hat{U}_\tau \rho_\mathrm{i}]+\frac{J}{N}\mathrm{Tr} [\hat{\sigma}_1^z \hat{\sigma}_N^z],
%\end{equation}
%where $\hat{h}_j (J,\Gamma)=-J \hat{\sigma}_j^z \hat{\sigma}_{j+1}^z-\Gamma \hat{\sigma}_j^x$.
%It is noted that the difference between $e_\mathrm{p}$
%For simplicity, we assume a periodic boundary condition.
%Here we omit a term $\Gamma \hat{\sigma}_N^x$ since it does not change the energy density in the thermodynamic limit.

The Lieb--Robinson bound~\cite{lieb1972finite} implies that $\hat{U}_\tau^\dagger \hat{h}_{j} (J, \{\Gamma\}) \hat{U}_\tau$ is effectively  supported within a region of radius $N_\mathrm{R} n_p\sim v \tau$ centered around site~$(j,k)$, where $v$ denotes the Lieb--Robinson velocity.
We denote this region by ${\mathcal S}_j=\{(j_\mathrm{min},1),\ldots,(j_\mathrm{max},n_p)\}$ and define the set of corresponding unit cells as ${\mathcal L}_j=\{j_\mathrm{min},\ldots, j_\mathrm{max} \}$, with $j_\mathrm{min}=j-N_\mathrm{R}+1$ and $j_\mathrm{max}=j+N_\mathrm{R}$.
For $N_\mathrm{R} n_p \ll N$ and $N_\mathrm{R}\leq j \leq \lfloor N/n_p \rfloor-N_\mathrm{R}$, Eq.~(\ref{eq:energyj}) becomes, up to $O(\epsilon)$ corrections with $\epsilon=\tau/N$,
\begin{equation}
    e_j=\mathrm{Tr} [\hat{U}_{\tau,j}^{\alpha\dagger} \hat{h}_{j}(J,\{\Gamma\}) \hat{U}_{\tau,j}^{\alpha} \rho_{\mathrm{i},j}]+O(\epsilon).
    \label{eq:local}
\end{equation}
Here, $\hat{U}^{\alpha}_{\tau,j}$ and $\rho_{\mathrm{i},j}$ act on ${\mathcal S}_j$, and $\alpha \in \{+1,-1\}$ specifies the boundary condition (periodic or antiperiodic).
Importantly, Eq.~(\ref{eq:local}) holds independently of the choice of boundary condition.
Explicitly,
\begin{equation}
    \left\{
    \begin{aligned}
    &\hat{U}_{\tau,j}^{\alpha}=\mathcal{T} e^{-i \int_0^\tau \hat{H}_j^{\alpha}\left(\frac{Jt}{\tau},\{\frac{\Gamma t}{\tau}+h_{(j,k)}\left(1-\frac{t}{\tau}\right)\} \right) dt}, \\
    &\rho_{\mathrm{i},j}=\bigotimes_{(\ell,k)\in {\mathcal S}_j} \Big[\cos^2 \phi_{(j,k)} \ket{+_{(\ell,k)}} \bra{+_{(\ell,k)}} \nonumber\\
    &\qquad \quad+\sin^2 \phi_{(j,k)} \ket{-_{(\ell,k)}} \bra{-_{(\ell,k)}}\Big].
    \end{aligned}
    \right.
\end{equation}
%where $0\leq \phi_{(j,k)} \leq \pi/4$ is ordered in descending order (see Eq.~(\ref{eq:condition})).
The Hamiltonian on ${\mathcal S}_j$ is defined as
\begin{align}
    &\hat{H}_j^{\alpha} (\tilde{J},\{\tilde{\Gamma}_k\})= \sum_{\ell \in \mathcal{L}_j} \hat{h}_{\ell} (\tilde{J},\{\tilde{\Gamma}_k\})\nonumber\\
    &\qquad\qquad\qquad+\tilde{J}\hat{\sigma}_{(j_\mathrm{max},n_p)}^z (\hat{\sigma}_{(j_\mathrm{max}+1,1)}^z-\alpha \hat{\sigma}_{(j_\mathrm{min},1)}^z).
    %\hat{H}^{(A)} (\tilde{J},\tilde{\Gamma};{\mathcal L})=& \sum_{j=j_\mathrm{min}}^{j_\mathrm{max}-1} \hat{h}_j (\tilde{J},\tilde{\Gamma})-\tilde{\Gamma} \hat{\sigma}_{j_\mathrm{max}}^x+\tilde{J}\hat{\sigma}_{j_\mathrm{max}}^z \hat{\sigma}_{1}^z.
\end{align}
%with the periodic boundary condition $\hat{\sigma}_{j_\mathrm{max}+1}^z=\hat{\sigma}_{j_\mathrm{min}}^z$.
%Here, we introduce the translationally invariant Hamiltonian and state, denoted by $\hat{H}_\mathrm{p}(J,\Gamma;\Lambda_j)$ and $\rho_{\mathrm{i},j}$, respectively.
Then, the local energy can be expressed as
\begin{equation}
    e_j=\sum_{\alpha\in\{\pm1\}}\mathrm{Tr} [\hat{U}_{\tau,j}^{\alpha\dagger} \hat{h}_j(J,\{\Gamma\}) \hat{U}_{\tau,j}^{\alpha} \rho_{\mathrm{i},j} \mathrm{P}_{-\alpha}] +O(\epsilon),
    %&+\mathrm{Tr} [\hat{U}_{\tau,j}^{(P)\dagger} \hat{h}_j(J,\Gamma) \hat{U}_{\tau,j}^{(P)}  \rho_{\mathrm{i},j} \mathrm{P}_-]+O(\epsilon)),
\end{equation}
where the projection operator $\mathrm{P}_{\alpha}$ is defined by
\begin{equation}
    \mathrm{P}_{\alpha}=\frac{1}{2} \left(1 +\alpha \prod_{i\in \mathcal{S}_j} \hat{\sigma}_i^x \right).
\end{equation}

Next, we introduce the translation operators $\hat{T}_j^{\alpha}$ for $\alpha \in \{+1, -1\}$, which act on operators in ${\mathcal S}_j$.
For $k \in \{1,\ldots,n_p \}$, they satisfy
\begin{equation}
    \left\{
    \begin{aligned}
    &\hat{T}_j^{\alpha\dagger} \hat{\sigma}_{(\ell,k)}^{x(z)} \hat{T}_j^\alpha= \hat{\sigma}_{(\ell+1,k)}^{x(z)} \text{ for } \ell \neq j_\mathrm{max},\\
    &\hat{T}_j^{\alpha \dagger} \hat{\sigma}_{(j_\mathrm{max},k)}^{x} \hat{T}_j^\alpha=\hat{\sigma}_{(j_\mathrm{min},k)}^{x},\\
    &\hat{T}_j^{\alpha \dagger} \hat{\sigma}_{(j_\mathrm{max},k)}^{z} \hat{T}_j^\alpha=\alpha \hat{\sigma}_{(j_\mathrm{min},k)}^{z},\\
    %&\hat{T}_j^{(A) \dagger} \hat{\sigma}_{j_\mathrm{max}}^{z} \hat{T}_j^A=-\hat{\sigma}_{j_\mathrm{min}}^{z}.
    \end{aligned}
    \right.
\end{equation}
Using these relations, we obtain for each $\alpha \in \{+1, -1\}$
\begin{align}
    &2N_\mathrm{R}\mathrm{Tr} [\hat{U}_{\tau,j}^{\alpha\dagger} \hat{h}_j(J,\{\Gamma\}) \hat{U}_{\tau,j}^{\alpha} \rho_{\mathrm{i},j} \mathrm{P}_{-\alpha}]\nonumber\\
    =&\sum_{n=1}^{2N_\mathrm{R}}\mathrm{Tr} [(\hat{T}_j^{\alpha \dagger})^{n}\hat{U}_{\tau,j}^{\alpha \dagger} \hat{h}_j(J,\{\Gamma\}) \hat{U}_{\tau,j}^{\alpha} \rho_{\mathrm{i},j} \mathrm{P}_{-\alpha} (\hat{T}_j^{\alpha})^{n}]\nonumber\\
    =&\mathrm{Tr} [\hat{U}_{\tau,j}^{\alpha\dagger} \hat{H}_j^{\alpha}(J,\{\Gamma\}) \hat{U}_{\tau,j}^{\alpha} \rho_{\mathrm{i},j} \mathrm{P}_{-\alpha} ],
\end{align}
%and
%\begin{align}
%    &2N_\mathrm{R}\mathrm{Tr} [\hat{U}_{\tau,j}^{(P)\dagger} \hat{h}_j(J,\Gamma) \hat{U}_{\tau,j}^{(P)} \rho_{\mathrm{i},j} \mathrm{P}_-]\nonumber\\
%    =&\sum_{n=1}^{2N_\mathrm{R}}\mathrm{Tr} [(\hat{T}_j^{(P) \dagger})^n\hat{U}_{\tau,j}^{(P)\dagger} \hat{h}_j(J,\Gamma) \hat{U}_{\tau,j}^{(P)} \rho_{\mathrm{i},j} \mathrm{P}_- (\hat{T}_j^{(P)})^n]\nonumber\\
%    =&\mathrm{Tr} [\hat{U}_{\tau,j}^{(P)\dagger} \hat{H}^{(P)}(J,\Gamma;\mathcal{L}_j) \hat{U}_{\tau,j}^{(P)} \rho_{\mathrm{i},j} \mathrm{P}_- ].
%\end{align}
%and the local energy is given by
which leads to
\begin{align}
    e_j=\frac{1}{2N_\mathrm{R}} \sum_{\alpha \in \{\pm 1\}}\mathrm{Tr} [\hat{U}_{\tau,j}^{\alpha\dagger} \hat{H}_j^{\alpha}(J,\{\Gamma\}) \hat{U}_{\tau,j}^{\alpha} \rho_{\mathrm{i},j} \mathrm{P}_{-\alpha}] +O(\epsilon).
    %&+\mathrm{Tr} [\hat{U}_{\tau,j}^{(P)\dagger} \hat{H}^{(P)}(J,\Gamma;\mathcal{L}_j) \hat{U}_{\tau,j}^{(P)}  \rho_{\mathrm{i},j} \mathrm{P}_-] \}+O(\epsilon),
\end{align}

Applying the Jordan--Wigner transformation, we obtain
\begin{equation}
    e_j=\frac{1}{2N_\mathrm{R}}\mathrm{Tr} [\hat{U}_{\tau,j}^{\dagger} \hat{H}_j(J,\{\Gamma\}) \hat{U}_{\tau,j} \rho_{\mathrm{i},j}] +O(\epsilon),
    \label{eq:appendix_ej}
\end{equation}
where
\begin{align}
    &\hat{H}_j(\tilde{J},\{\tilde{\Gamma}_k\})\nonumber\\
    =&-\tilde{J} \Big[\sum_{i \in \mathcal{S}_j \setminus (j_\mathrm{max},n_p)} (\hat{c}_{i}^\dagger-\hat{c}_{i})(\hat{c}_{i+1}^\dagger +\hat{c}_{i+1}) \nonumber\\
    %&-(\hat{c}_{(j_\mathrm{max},n_p)}^\dagger -\hat{c}_{(j_\mathrm{max},n_p)})(\hat{c}_{{(j_\mathrm{max}+1,1)}}^\dagger+\hat{c}_{{(j_\mathrm{max}+1,1)}}) \nonumber\\
    &+ (\hat{c}_{(j_\mathrm{max},n_p)}^\dagger -\hat{c}_{(j_\mathrm{max},n_p)})(\hat{c}_{{(j_\mathrm{min},1)}}^\dagger+\hat{c}_{{(j_\mathrm{min},1)}} \Big]\nonumber\\
    &-\sum_{(\ell,k) \in \mathcal{S}_j} \tilde{\Gamma}_k (2\hat{c}_{(\ell,k)}^\dagger \hat{c}_{(\ell,k)} -1)\nonumber\\
    =&\sum_{i \in\mathcal{S}_j} \Lambda_{i}^{(\tilde{J},\{\tilde{\Gamma}_k\})} \left(\hat{\eta}_{i}^{(\tilde{J},\{\tilde{\Gamma}_k\})\dagger} \hat{\eta}_{i}^{(\tilde{J},\{\tilde{\Gamma}_k\})} -\frac{1}{2} \right)
    \label{eq:period}
\end{align}
with fermionic operators $\hat{\eta}_{i}^{(\tilde{J},\{\tilde{\Gamma}_k\})}$ and $\hat{\eta}_{i}^{(\tilde{J},\{\tilde{\Gamma}_k\})\dagger}$ and single-particle energies $\Lambda_{i}^{(\tilde{J},\{\tilde{\Gamma}_k\})}$, ordered in ascending order.
The time-evolution operator reads
\begin{equation}
    \hat{U}_{\tau,j}=\mathcal{T}e^{-i \int_0^\tau \hat{H}_j \left( \frac{Jt}{\tau}, \{\frac{\Gamma t}{\tau}+h_{(j,k)}\left( 1-\frac{t}{\tau} \right)\} \right) dt}.
    \label{eq:Uj}
\end{equation}
%Here, $\hat{\eta}_{i}^{(\tilde{J},\{\tilde{\Gamma}_k\})}$ and $\hat{\eta}_{i}^{(\tilde{J},\{\tilde{\Gamma}_k\})\dagger}$ are fermionic annihilation and creation operators, and $\Lambda_{i}^{(\tilde{J},\{\tilde{\Gamma}_k\})}$ denotes the single particle energies, ordered in ascending order.
The energy can be expressed as
\begin{align}
    &\mathrm{Tr} [\hat{U}_{\tau,j}^{\dagger} \hat{H}_j(J,\{\Gamma\}) \hat{U}_{\tau,j} \rho_{\mathrm{i},j}]\nonumber\\
    %=&\sum_{(\ell,k), (\ell',k') \in \mathcal{S}_j} \frac{\Lambda_{(\ell,k)}^{(J,\{\Gamma\})}}{2} (|u_{(\ell,k),(\ell',k')}|^2-|v_{(\ell,k),(\ell',k')}|^2) \cos 2\phi_{(j,k')},\nonumber\\
    =&\sum_{(\ell,k), (\ell',k') \in \mathcal{S}_j} \frac{\Lambda_{(\ell,k)}^{(J,\{\Gamma\})}}{2} A_{(\ell,k),(\ell',k')} \cos 2\phi_{(j,k')},
\end{align}
where $A_{i,i'}=|u_{i,i'}|^2-|v_{i,i'}|^2$ and $\hat{U}_{\tau,j}^{\dagger} \hat{\eta}_{i}^{(J,\{ \Gamma\})} \hat{U}_{\tau,j} = \sum_{i' \in \mathcal{S}_j} (u_{i,i'} \hat{c}_{i'}+v_{i,i'}^* \hat{c}_{i'}^\dagger)$.
The coefficients satisfy $\sum_{i'' \in \mathcal{S}_j} (u_{i,i''}u_{i',i''}^*+v_{i,i''}^* v_{i',i''})=\delta_{i,i'}$.

We now derive a lower bound for the energy.
Using the von-Neumann's trace inequality~\cite{mirsky1975trace}, we obtain
\begin{align}
    &\mathrm{Tr} [\hat{U}_{\tau,j}^{\dagger} \hat{H}_j(J,\{\Gamma\}) \hat{U}_{\tau,j} \rho_{\mathrm{i},j}]\nonumber\\
    \geq&-\frac{1}{2}\sum_{(\ell,k), (\ell',k') \in \mathcal{S}_j} \Lambda_{(\ell,k)}^{(J,\{\Gamma\})} |v_{(\ell,k),(\ell',k')}|^2 \cos 2\phi_{(j,k')} \nonumber\\
    \geq&-\frac{1}{2} \sum_{(\ell,k) \in \mathcal{S}_j} \Lambda_{[k,\ell]}^{(J,\{\Gamma\})} \cos 2\phi_{(j,k)},
    \label{eq:appendix_inequality}
\end{align}
where $[k,\ell]=2(k-1)N_\mathrm{R}+\ell +(j_\mathrm{min}-1)(n_p-1)$.
%where we have used 
%The equality is achieved when $\hat{U}_{\tau,j}^{\dagger} \hat{\eta}_{(\ell,k)}^{(J,\{ \Gamma\})} \hat{U}_{\tau,j} = \sum_{\ell' \in \mathcal{L}_j} v_{\tau,(\ell,k),(\ell',k)}^* \hat{c}_{(\ell',k)}^\dagger$.
Consequently, Eqs.~(\ref{eq:appendix_ej}) and~(\ref{eq:appendix_inequality}) yield
\begin{align}
    &e(\rho_\mathrm{f}(\tau))=\frac{1}{N} \sum_{j=N_\mathrm{R}}^{\lfloor N/n_p \rfloor -N_\mathrm{R}} e_j +O(\epsilon) \nonumber\\
    &\geq -\sum_{j=N_\mathrm{R}}^{\lfloor N/n_p \rfloor -N_\mathrm{R}} \sum_{(\ell,k) \in \mathcal{S}_j} \frac{\Lambda_{[k,\ell]}^{(J,\{\Gamma \})} \cos (2\phi_{(j,k)})}{4NN_R} +O(\epsilon).
\end{align}
Taking first $N\to\infty$ and subsequently $\tau \to \infty$, and using the relation given in Eq.~(\ref{eq:condition}), we arrive at Eq.~(\ref{eq:bound}).

We now discuss the tightness of the bound.
For $n_p=1$, following the argument leading to Eq.~(\ref{eq:relation_QA}), we obtain the equality, 
\begin{equation}
    \mathrm{Tr} [\hat{U}_{\tau,j}^{\dagger} \hat{H}_j(J,\{\Gamma\}) \hat{U}_{\tau,j} \rho_{\mathrm{i},j}]
    = 2N_\mathrm{R} e'_\mathrm{a} (\tau) \cos (2\phi_j),
    \label{eq:appendix_final}
\end{equation}
where $e'_\mathrm{a} (\tau)$ denotes the energy density of $\hat{H}_j(J,\{\Gamma\})$ after the quantum annealing process.
In the thermodynamic limit, the bound in Eq.~(\ref{eq:bound}) is achieved provided that $e'_\mathrm{a}(\tau)$ converges to the ground-state energy density $e'_\mathrm{g.s.}$.
Adiabatic convergence to the ground state is guaranteed only in the absence of a quantum phase transition.
When a continuous quantum phase transition is crossed, excitations are generated according to the Kibble--Zurek mechanism.
Nevertheless, the excess energy density vanishes algebraically as $\tau\to\infty$ (for the present model, $e'_\mathrm{a}(\tau)-e'_\mathrm{g.s.} \sim \tau^{-1/2}$), implying that the ground-state energy density, and hence the bound in Eq.~(\ref{eq:bound}), is asymptotically recovered as $\tau \to \infty$, even though the final state remains excited.
%implying that the ground-state energy density is recovered in the infinite-time limit even though the final state remains excited.
%$e'_\mathrm{a}(\tau) - e'_\mathrm{a}(\infty) \sim \tau^{-\nu/(1+z\nu)}$, where $e'_\mathrm{a}(\infty)$ coincides with the ground-state energy and, $z$ and $\nu$ are the dynamical and correlation-length critical exponents, respectively.
%For the present model, when the parameters $(J, \Gamma)$ lie in the ferromagnetic phase, the dynamics passes through a quantum critical point with $z=1$ and $\nu=1$, yielding $e'_\mathrm{a}(\tau)-e'_\mathrm{a}(\infty) \sim \tau^{-1/2}$.

For finite $n_p\geq 2$, the bound in Eq.~(\ref{eq:appendix_inequality}) is saturated provided that the adiabatic condition $u_{i,i'}=0$ for $i,i'\in \mathcal{S}_j$ and $v_{[k,\ell],(\ell,k')}=0$ for $k\neq k'$ is satisfied.
%the Hamiltonian in Eq.~(\ref{eq:Uj}) can be reduced, after Fourier transformation, to a Bogoliubov--de Gennes matrix of $2n_p$ dimensions.
%As a consequence, the low-energy spectrum contains a linearly dispersing mode, and
Two mechanisms can violate this condition in the thermodynamic limit: the presence of a quantum phase transition and inter-band transitions between different $k$.
Concerning the former, the Hamiltonian $\hat{H}_j(\tilde{J},\{\Gamma_k\})$ belongs to the same universality class as the uniform case (i.e., $n_p=1$), as shown in Appendix~\ref{appendix:universality}.
The Kibble--Zurek mechanism therefore implies that the associated excess energy density decays as $\tau^{-1/2}$ and hence vanishes in the limit $\tau \to \infty$.
In addition, for $n_p \geq 2$, the inter-band gap closes with increasing $\tau$, which violates the assumptions of the adiabatic theorem.
Although a rigorous proof is still lacking, the numerical results presented in the main text indicate that inter-band transitions become irrelevant at the level of energy density (equivalently, specific relative entropy) in the limit $\tau \to \infty$.
Consequently, we conclude that the bound in Eq.~(\ref{eq:bound}) is asymptotically saturated as $\tau \to \infty$ for all finite $n_p$ considered in this work.

\section{Universality class of periodically modulated systems}\label{appendix:universality}
This appendix demonstrates that the Hamiltonian in Eq.~(\ref{eq:period}) with finite $n_p$ belongs to the same universality class as the uniform transverse-field Ising model ($n_p=1$).
In the following, we consider $n_p\geq 2$.

After Fourier transformation, the Hamiltonian is reduced to a Bogoliubov--de Gennes (BdG) Hamiltonian $\mathcal{H}_j(k_n)$ of dimension $2n_p$:
\begin{equation}
    \hat{H}_j (\tilde{J},\{\tilde{\Gamma}_k\})=-\frac{1}{2}\sum_{n=0}^{N_\mathrm{R}-1} \Phi_n^\dagger \mathcal{H}_j(k_n) \Phi_n,
\end{equation}
where $\Phi_n=(c_{n,1},\ldots,c_{n,n_p},c_{-n,1}^\dagger,\ldots,c_{-n,n_p}^\dagger)^\top$ and $k_n=2\pi n/N_\mathrm{R}$ for $n\in \mathbb{Z}$.
%Here, $\{\hat{c}_{n,j}\}_{j=1}^{n_p}= \sum_{\ell=1}^{n_p} c_{(\ell,j)}e^{-ik_n\ell}/N_\mathrm{R}$ is a fermion operator. 
%Here, $k_n=2\pi n/N_\mathrm{R}$ for $n\in \mathbb{Z}$ denotes the momentum, and $(\cdot)^\top$ denotes the transpose.
%the Fourier transformation introduces fermion operators $\{\hat{c}_{n,j}\}_{j=1}^{n_p}$ and $\{\hat{c}_{k,j}^\dagger\}_{j=1}^{n_p}$ as
%\begin{equation}
%    \left\{
%    \begin{aligned}
%    c_{n,j}=\frac{1}{N_\mathrm{R}} \sum_{\ell=1}^{n_p} c_{(\ell,j)}e^{-ik_n\ell}\\
%    c_{n,j}^\dagger=\frac{1}{N_\mathrm{R}} \sum_{\ell=1}^{n_p} c_{(\ell,j)}^\dagger e^{-ik_n\ell},
%    \end{aligned}
%    \right.
%\end{equation}
%where $k_n=2\pi n/N_\mathrm{R}$ and $n\in \mathbb{R}$.
%Then, the Hamiltonian can be written as $\hat{H}_j (\tilde{J},\{\tilde{\Gamma}_k\})=-\sum_{n=0}^{N_\mathrm{R}-1} \Phi_n^\dagger \mathcal{H}_j(k_n) \Phi_n/2$, where $\Phi_n=(c_{n,1},\ldots,c_{n,n_p},c_{-n,1}^\dagger,\ldots,c_{-n,n_p}^\dagger)^\top$ and $(\cdot)^\top$ denotes the transpose.
The BdG Hamiltonian can be written in block form as
\begin{equation}
    \mathcal{H}_j(k)=
    \begin{pmatrix}
    A_j(k) & B_j(k)\\
    -B_j(k) & -A_j(k)
    \end{pmatrix}
\end{equation}
with $n_p$-dimensional matrices $A_j(k)$ and $B_j(k)$.
The matrix elements are given by
\begin{equation}
    [A_j(k)]_{\ell,m}=\left\{
    \begin{aligned}
        &2\tilde{\Gamma}_\ell \text{ for } \ell=m\\
        &\tilde{J} \text{ for } m=\ell\pm 1\\
        &\tilde{J} e^{-ik} \text{ for } (\ell,m)=(1, n_p)\\
        &\tilde{J} e^{ik} \text{ for } (\ell,m)=(n_p, 1)\\
        &0 \text{ otherwise,}
    \end{aligned}
    \right.
\end{equation}
and
\begin{equation}
    [B_j(k)]_{\ell,m}=\left\{
    \begin{aligned}
        &2\tilde{\Gamma}_\ell \text{ for } \ell=m\\
        &\tilde{J} \text{ for } m=\ell+ 1\\
        &-\tilde{J} \text{ for } m=\ell - 1\\
        &-\tilde{J} e^{-ik} \text{ for } (\ell,m)=(1, n_p)\\
        &\tilde{J} e^{ik} \text{ for } (\ell, m)=(n_p, 1)\\
        &0 \text{ otherwise,}
    \end{aligned}
    \right.
\end{equation}
where $[\cdot]_{\ell,m}$ is the $(\ell,m)$ matrix element.
The BdG Hamiltonian belongs to the BDI symmetry class, possessing particle-hole symmetry, spinless time-reversal symmetry, and chiral symmetry~\cite{schnyder2008classification}.
Owing to the chiral symmetry, $\mathcal{H}_j(k)$ can be transformed into an off-diagonal form by a unitary transformation:
\begin{equation}
    \tilde{\mathcal{H}}_j(k)=
    \begin{pmatrix}
    0 & C_j(k)\\
    C_j^\dagger(k) & 0
    \end{pmatrix},
\end{equation}
where $C_j(k)$ is an $n_p$-dimensional matrix with matrix elements given by
\begin{equation}
    [C_j(k)]_{\ell,m}=\left\{
    \begin{aligned}
        &2 \tilde{\Gamma}_\ell \text{ for } \ell=m\\
        &2 \tilde{J} \text{ for } \ell-m=1\\
        &2\tilde{J} e^{-ik} \text{ for } \ell=1, m=n_p\\
        &0 \text{ otherwise.}
    \end{aligned}
    \right.
\end{equation}

In the thermodynamic limit, $k \in (-\pi,\pi]$.
The excitation spectrum is given by $\pm|e_n(C_j (k))|$, where $e_n (C)$ denotes the $n$-th eigenvalue of $C$.
%Now, we evaluate the critical point.
%
%The eigenmode of $\tilde{\mathcal{H}}_j(k)$ is $\{ \pm |\mathrm{e}_n(C_j (k))| \}_{n=1}^{n_p}$, where $e_n (C)$ is the $n$-th eigenvalue of matrix $C$.
Therefore, the quantum critical point is determined by the condition $\det C_j(k)=0$ and explicitly given by
\begin{equation}
    g\coloneqq \tilde{J}-\left(\prod_{\ell=1}^{n_p}\tilde{\Gamma}_\ell \right)^{1/n_p} =0,
    %=2^{n_p}\left(\prod_{n=1}^{n_p}\Gamma_n +(-1)^{n_p}J^{n_p}e^{-ik}\right)=0.
\end{equation}
with $k=0$ for even $n_p$ and $k=\pi$ for odd $n_p$.

Let $e_1(C_j(k))$ be the eigenmode that becomes gapless at the critical point and $k_\mathrm{c}$ be the corresponding momentum.
Expanding around $k=k_\mathrm{c}$ and $g=0$, we obtain $e_1(C_j(k))\simeq v (k-k_\mathrm{c})+ag$ with complex constants $v$ and $a$.
Since the BdG Hamiltonian belongs to the BDI class, the spectrum is symmetric under $k \to -k$.
Consequently, the excitation gap behaves as
\begin{equation}
    |e_1(C_j(k))|\simeq \sqrt{|v(k-k_c)|^2+|ag|^2}.
\end{equation}
Therefore, the critical exponents are $z=1$ and $\nu=1$, identical to those of the uniform transverse-field Ising model.

\bibliography{tvqe}

\end{document}